\documentclass[aps,prd,10pt,twocolumn,superscriptaddress,showpacs]{revtex4-1}
\pdfoutput=1
\usepackage{hyperref}
\usepackage{graphicx}
\usepackage{enumerate}
\usepackage{amsfonts,amsmath,amssymb,bm,bbm}
\usepackage{color}
\usepackage{epsfig, subfigure}
\usepackage{setspace}
\usepackage{booktabs, tabularx} 
\usepackage{float}

\usepackage{amssymb}
\usepackage{amsmath}
\usepackage{url}
\usepackage{booktabs}

\hypersetup{
    pdfnewwindow=true,      
    colorlinks=true,       
    linkcolor=blue,          
    citecolor=blue,        
    filecolor=blue,      
    urlcolor=blue        
}

\newcommand{\be}{\begin{equation}}  
\newcommand{\ee}{\end{equation}}
\newcommand{\bea}{\begin{eqnarray}}  
\newcommand{\eea}{\end{eqnarray}}

\begin{document}

\title{\mbox{\hspace{-0.25cm} Resolving Small-Scale Dark Matter Structures Using Multi-Source Indirect Detection }}

\author{Kenny C. Y. Ng}
\affiliation{Center for Cosmology and AstroParticle Physics (CCAPP), Ohio State University, Columbus, OH 43210}
\affiliation{Department of Physics, Ohio State University, Columbus, OH 43210}

\author{Ranjan Laha}
\affiliation{Center for Cosmology and AstroParticle Physics (CCAPP), Ohio State University, Columbus, OH 43210}
\affiliation{Department of Physics, Ohio State University, Columbus, OH 43210}

\author{Sheldon Campbell}
\affiliation{Center for Cosmology and AstroParticle Physics (CCAPP), Ohio State University, Columbus, OH 43210}
\affiliation{Department of Physics, Ohio State University, Columbus, OH 43210}

\author{\mbox{Shunsaku Horiuchi}}
\affiliation{Center for Cosmology, Department of Physics and Astronomy, University of California, Irvine, CA 92697}

\author{Basudeb Dasgupta}
\affiliation{International Center for Theoretical Physics, 34014 Trieste, Italy}

\author{Kohta Murase}
\affiliation{Hubble Fellow, School of Natural Sciences, Institute for Advanced Study, Princeton, NJ 08540}

\author{John F. Beacom}
\affiliation{Center for Cosmology and AstroParticle Physics (CCAPP), Ohio State University, Columbus, OH 43210}
\affiliation{Department of Physics, Ohio State University, Columbus, OH 43210}
\affiliation{Department of Astronomy, Ohio State University, Columbus, OH 43210 \\ 
{\tt ng.199@osu.edu, laha.1@osu.edu, campbell.1431@osu.edu, s.horiuchi@uci.edu, bdasgupta@ictp.it, murase@ias.edu, beacom.7@osu.edu} \smallskip}
\date{October 14, 2013}

\begin{abstract}

The extragalactic dark matter (DM) annihilation signal depends on the product of the clumping factor, $\langle \delta^{2} \rangle $, and the velocity-weighted annihilation cross section, $\sigma v$. This ``clumping factor--$\sigma v$'' degeneracy can be broken by comparing DM annihilation signals from multiple sources. In particular, one can constrain the minimum DM halo mass, $M_{\rm min}$, which depends on the mass of the DM particles and the kinetic decoupling temperature, by comparing observations of individual DM sources to the diffuse DM annihilation signal. We demonstrate this with careful semi-analytic treatments of the DM contribution to the diffuse Isotropic Gamma-Ray Background (IGRB), and compare it with two recent hints of DM from the Galactic Center, namely, $\sim130$ GeV DM annihilating dominantly in the $\chi\chi \to \gamma \gamma$ channel, and $(10-30)$ GeV DM annihilating in the $\chi\chi \to b\bar{b}$ or $\chi\chi \to \tau^{+}\tau^{-}$ channels. We show that, even in the most conservative analysis, the Fermi IGRB measurement already provides interesting sensitivity. A more detailed analysis of the IGRB, with new Fermi IGRB measurements and modeling of astrophysical backgrounds, may be able to probe values of $ M_{\rm min}$ up to $\sim 1\,{ M_{\odot}}$ for the 130 GeV candidate and $\sim 10^{-6}\,{ M_{\odot}}$ for the light DM candidates. Increasing the substructure content of halos by a reasonable amount would further improve these constraints.

\end{abstract}



\maketitle

\section{Introduction}
\label{sec:introduction}

The observed universe is well explained by the $\rm{\Lambda CDM} $ cosmological model ($\rm{\Lambda} $ Cold Dark Matter). A large fraction, $\Omega_{\Lambda} $, of its energy density is in the form of enigmatic dark energy, and the rest, $\Omega_{M} $,  is mostly non-relativistic matter and a tiny fraction of relativistic particles. A major fraction of $\Omega_{M}$ has no detectable electromagnetic interactions, thus is termed Dark Matter (DM). From its gravitational effects on different length scales, DM is determined to have an energy density fraction $\Omega_{\chi}$. The particle nature of DM is largely unknown. 

Identifying the fundamental particle nature of DM is one of the most important problems in contemporary science. A well-motivated DM candidate is the generic Weakly Interacting Massive Particle (WIMP), produced as a thermal relic in the early universe \cite{Jungman:1995df, Bergstrom:2000pn, Bertone:2004pz}. DM that self-annihilates at the electroweak scale naturally produces the observed DM abundance. The precise value of the thermally averaged total annihilation cross section that determines the DM abundance depends on several parameters \cite{Steigman:2012nb}. A larger value will delay chemical decoupling, which would underproduce DM relative to the observed abundance, and vice versa. 

After freeze-out, DM will continue to self-annihilate but at a cosmologically negligible rate. At the present epoch, DM is non-relativistic and is no longer thermally distributed. As a result, the velocity-weighted cross section (or simply annihilation cross section), $\sigma v$, which controls the annihilation rate now, could be a function of relative velocity and thus depends on the phase space of the DM structures. In this work, we consider the simplest case where $\sigma v$ is velocity independent over the relevant range of velocities of cosmic DM. In this case (s-wave), if the total value of $\sigma v $ now differs from the value of thermally-averaged $\langle\sigma v \rangle$ that determines the relic abundance, it could imply a dominantly p-wave annihilation cross section \cite{Campbell:2010xc,Campbell:2011kf}, non-trivial velocity dependence of $\sigma v $ \cite{Hisano:2004ds, ArkaniHamed:2008qn}, or some special thermal scenarios \cite{Griest:1990kh}.

DM self-annihilation opens up the possibility of remotely detecting its annihilation products from concentrated DM sources, i.e., indirect detection. Together with directly detecting nuclear recoils in underground experiments, DM production in collider experiments, and DM influence of astrophysical systems, these four types of DM detection provide crucial and complementary information on the particle nature of DM \cite{Strigari:2012gn}. 

Indirect detection is a powerful way to detect DM. However, it suffers from problems of low signal-to-noise ratios due to large and complicated astrophysical backgrounds. One strategy is to search for smoking-gun signatures that would allow for effective separation between background and signal. Examples of such signatures are spectral lines \cite{Bergstrom:1988fp, Bergstrom:1997fj, Gustafsson:2007pc, Bertone:2009cb}, spectral cut-offs \cite{Beacom:2006tt, Bringmann:2007nk, Mack:2008wu, Bergstrom:2008gr, Ibarra:2012dw}, or distinct anisotropy signals \cite{Ando:2005xg, SiegalGaskins:2008ge, SiegalGaskins:2009ux}. Since annihilation signals are proportional to the DM density squared \cite{Gunn:1978gr, Zeldovich:1980st}, it is advantageous to search for these signatures from regions where DM is clustered, e.g., the Galactic Center (GC), dwarf galaxies, galaxy halos, galaxy clusters, or the diffuse signal from annihilation in all the DM structures in the Universe. 

The diffuse extragalactic DM annihilation signal is particularly difficult to predict robustly \cite{Bergstrom:2001jj, Ullio:2002pj, Taylor:2002zd}. It depends not only on the self-annihilation cross section and DM density distribution within halos, but also on the statistics of cosmological DM halos such as the halo abundances and their concentrations at small scales. Dense DM structures are expected to be present, whether they are isolated or residing within halos as substructures. They span down to the smallest possible bound DM objects with mass $ M_{\rm min}$. 

The value of $ M_{\rm min}$ corresponds to the cutoff of the matter power spectrum, $\rm k_{max}$, which is usually set by either the free-streaming scale after kinetic decoupling \cite{Green:2005fa} or the scale of acoustic oscillation with the radiation fields \cite{Loeb:2005pm}. These scales depend on the DM mass and its elastic coupling to the cosmic background particles. Parameter scans of some supersymmetric models show a large range of possibilities, $10^{-12}\,{ M_{\odot}}<{M_{\rm min}}<10^{-3}\,{M_{\odot}} $ \cite{Profumo:2006bv, Bringmann:2009vf}.

Direct observation of microhalos is very difficult because they are not massive enough to host stars. Current gravitational lensing probes are only sensitive to relatively massive halos ($>  10^{6}\, M_{\odot}$) \cite{Dalal:2001fq,2009MNRAS.398.1235X, 2010MNRAS.408.1969V}.  Nanolensing \cite{Chen:2010ae, Garsden:2011bz} or proper motion detection \cite{Koushiappas:2006qq,GeringerSameth:2010hm} might be able to prober smaller scales. The presence of microhalos, however, changes the clustering property of DM structures, which is encoded in the clumping factor, $\langle\delta^{2}(z) \rangle $ \cite{Bergstrom:2001jj, Ullio:2002pj, Taylor:2002zd, Serpico:2011in}, defined as the mean of the matter overdensity squared. The clumping factor boosts the annihilation rate relative to the mean background density, and is completely degenerate with the effect of $\sigma v$ for extragalactic diffuse DM annihilation signals. Both the annihilation cross section and the clumping factor are important DM parameters to be determined. 

In this work, we demonstrate how to break this degeneracy and constrain both ${ M_{\rm min}}$ and $\sigma v$ by comparing the diffuse Isotropic Gamma-ray Background (IGRB) \cite{Abdo:2010nz, Abdo:2010dk} with tentative DM annihilation signals from the GC \cite{Bringmann:2012vr,Weniger:2012tx, Hooper:2011ti, Abazajian:2012pn}. These excesses of events from GC might be DM signals, astrophysical phenomena, or experimental artifacts. It is important to scrutinize them as much as possible. We therefore consider them as a proof of principle as well as a test. Multiple-source analyses for DM indirect detection have proven to be invaluable for constraining DM candidate signals \cite{Ando:2005hr, Pinzke:2009cp}. 

Throughout this work, we use ${ M}_{x} = M/10^{x}\, { M_{\odot}} $ and cosmological parameters from the Planck mission ($\Omega_{\Lambda}=0.6825$, $\Omega_{M} =0.3175$, $\Omega_{\chi}= 0.1203\, h^{-2}$, $h=0.6711$, and the Hubble constant, $H_{0} = 100\,h\, {\rm km\,s^{-1}\,Mpc^{-1}}$, $n_{s} = 0.96$, $\sigma_{8}=0.8344$)  \cite{Ade:2013zuv}.

In Sec.~\ref{sec:DM}, we calculate the contribution of DM annihilation signals in the IGRB, showing the dependence of annihilation signals on $ M_{\rm min}$. In Sec.~\ref{sec:constrains}, we discuss the constraints obtained by combining the GC and IGRB observations for DM candidate events. Lastly, we summarize in Sec.~\ref{sec:summary}. 

\section{IGRB from DM annihilation}
\label{sec:DM}

The diffuse IGRB is the isotropic component of the gamma-ray sky, in principle mostly contributed by unresolved extragalactic astrophysical sources. The Fermi Gamma-Ray Space Telescope measures the IGRB by careful reductions of the Galactic astrophysical components, astrophysical sources, and detector backgrounds \cite{Abdo:2010nz}. In the presence of DM annihilation, it contains an irreducible isotropic Galactic component and the diffuse extragalactic component \cite{Yuksel:2007ac,Abazajian:2010sq}. In this section, we discuss each component and their dependence on DM substructures. 

\subsection{Isotropic Galactic component}

Since Fermi is embedded in the Milky Way (MW), an irreducible isotropic contribution of DM self-annihilation to the IGRB comes from the MW halo. We first review the case of DM annihilation in the MW.

The smooth DM density distribution in the MW, $\rho_{\chi}^{\alpha\beta\gamma}(r)$, can be parametrized by the following form \cite{Bringmann:2012ez}, 
\begin{align}
  \rho^{\alpha\beta\gamma}_{\chi}(r) = \rho_\odot \left[ \frac{r}{r_\odot}
  \right]^{-\gamma} \left[
  \frac{1+(r_\odot/r_s)^\alpha}{1+(r/r_s)^\alpha}
  \right]^{\frac{\beta-\gamma}{\alpha}}\,,
  \label{eqn:abgProfile}
\end{align}
where $r$ is the galactocentric distance, $\rho_{\odot}$ = $ 0.4\pm 0.1\, {\rm GeV\,cm^{-3}}$ is the DM density in the solar neighborhood \cite{ Bovy:2012tw, Salucci:2010qr}, $r_{\odot} = 8.5\, {\rm kpc}$ is the solar distance to the GC, and $r_{s}$ is the scale radius. The shape of the profile is determined by the parameters, $\alpha$, $\beta$, $\gamma$, and the scale radius, $r_{s}$. The commonly used NFW profile in the MW takes the values $\{\alpha, \beta, \gamma, r_{s}\} = \{1, 3, 1, 20\,{\rm kpc}\}$; the cored isothermal (ISO) profile takes  $\{2, 2, 0, 3.5\,{\rm kpc}\}$. Another profile favored by recent simulations is the Einasto (EIN) profile \cite{Diemand:2008in}, 
\begin{equation}
\rho^{Ein}_{\chi}(r) = \rho_{\odot}\, {\rm exp}\left( -\frac{2}{\alpha_{E}} \frac{r^{\alpha_{E}} - r_{\odot}^{\alpha_{E}}}{r_{s}^{\alpha_{E}}} \right)\, ,
\end{equation}
with $\alpha_{E} = 0.17 $ and $r_{s} = 20\, {\rm kpc}$. 

The gamma-ray (number flux) intensity due to the Galactic Halo DM self annihilation, $I_{\gamma}^{G}(E_{0})$, is 
\begin{eqnarray}
\nonumber I_{\gamma}^{\rm G}(E_{0}) &=& \frac{dN_{\gamma}}{dA\,dt_{0}\,d\Omega\, dE_{0}} \\ 
&=& \frac{{\sigma v} }{8\pi} \frac{r_{\odot}\rho_{\odot}^{2}}{m_{\chi}^{2}} {\cal J}(\psi) \frac{dN_{\gamma}}{dE_{0}} \\
\nonumber &=& 3.7\times 10^{-10}\, {\rm cm^{-2}\,s^{-1}\,sr^{-1}\,GeV^{-1}} \times{\cal J}(\psi)  \\
\nonumber && \left[ \frac{\sigma v}{2.2\times 10^{-26} \, {\rm cm^{3}\,s^{-1}} } \right] \left[ \frac{100\,{\rm GeV}}{m_{\chi}} \right]^{2} \frac{dN_{\gamma}}{dE_{0}/{\rm GeV}}\, ,
\end{eqnarray}
where $m_{\chi}$ is the DM mass, $E_{0}$ is the observed photon energy, and ${dN_{\gamma}}/{dE_0}$ is the photon energy spectrum per annihilation. The so-called J-factor, ${\cal J}(\psi)$, is the dimensionless line of sight integral of the density squared, and depends on the DM distribution in the Galactic halo, including halo substructures. 

\begin{figure}[t]
\includegraphics[angle=0.0, width=\columnwidth]{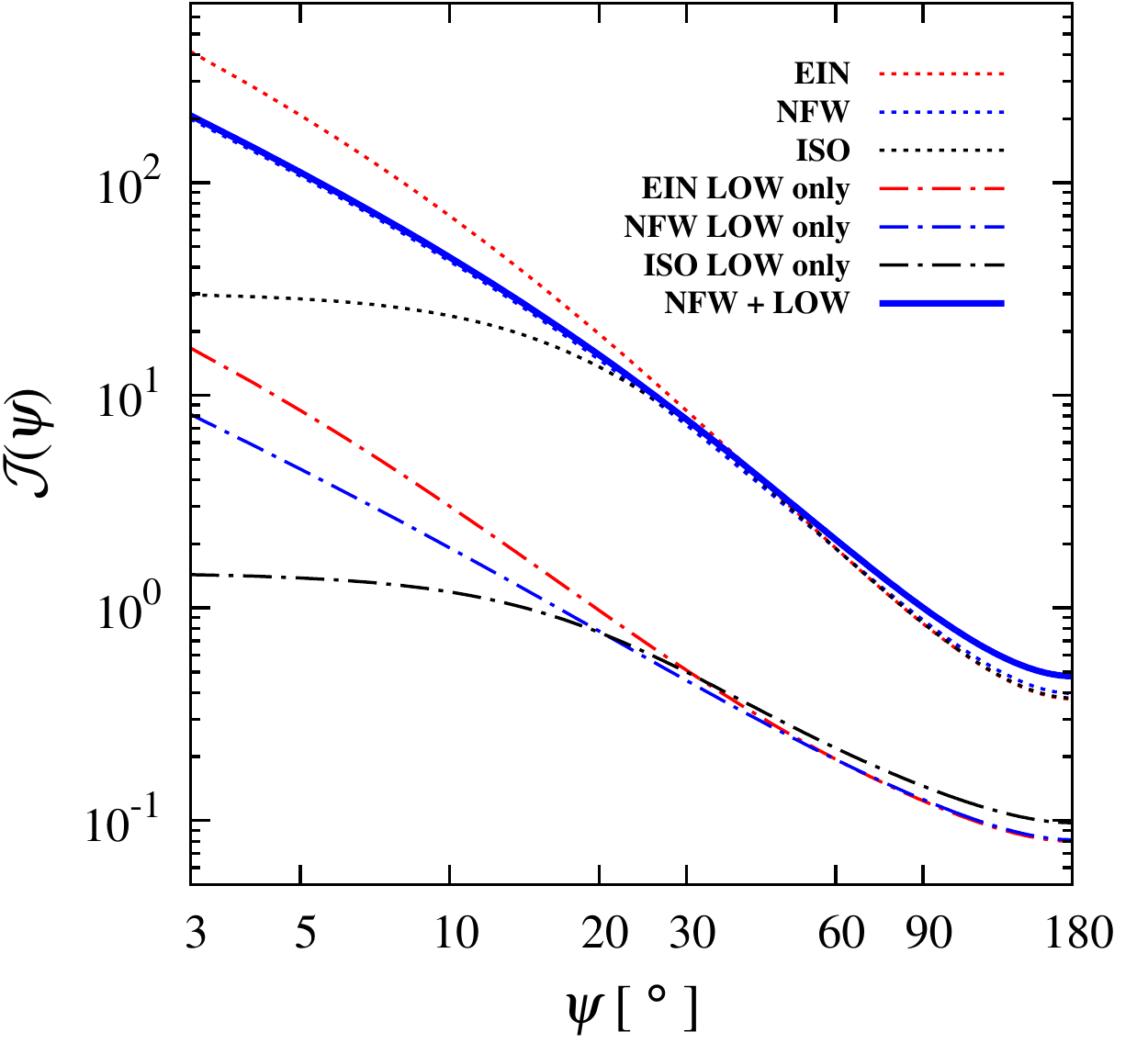}
\caption{The normalized line of sight integral of the DM density squared (the J-factor) as a function of the viewing angle, $\psi$. The J-factor for EIN, NFW and ISO profiles for the smooth halo are shown with dotted lines. The contributions of of substructures to the J-factors for the LOW substructure case, assuming ${ M_{\rm min}} = 10^{-6} \,{ M_{\odot}}$, are shown with dotted-dashed lines. The total J-factor (smooth + LOW substructure) for just the one case (NFW), is shown with a bold solid line. 
} 
\label{fig:plot_jfactor_LOW}
\end{figure}

\subsubsection{Galactic smooth halo}

\begin{figure}[t]
\includegraphics[angle=0.0, width=\columnwidth]{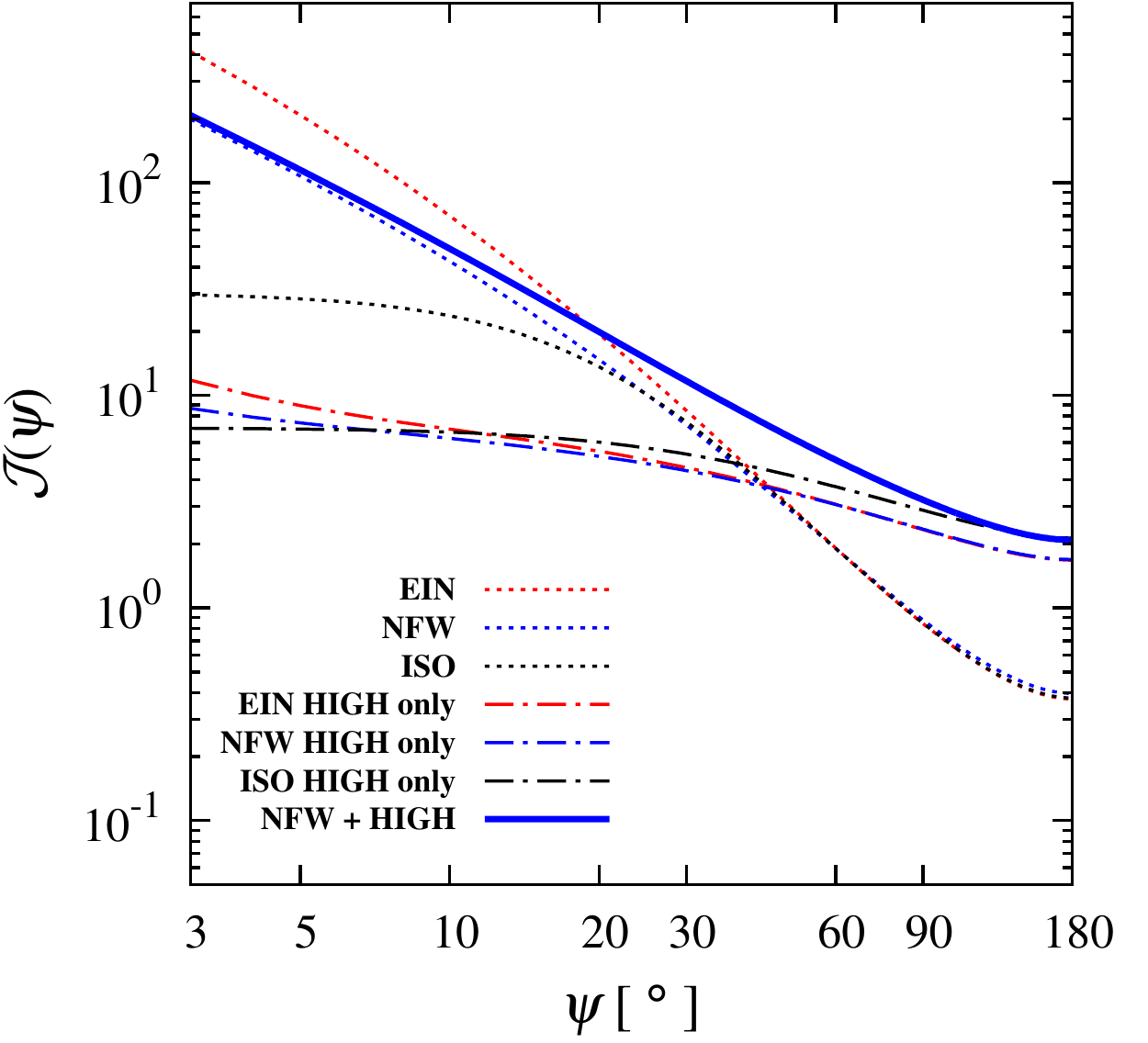}
\caption{Same as Fig.~\ref{fig:plot_jfactor_LOW}, but for the HIGH substructure case. 
} 
\label{fig:plot_jfactor_HIGH}
\end{figure}

The J-factor for the smooth DM density distribution for an observer within the halo, as a function of the angle between the line of sight and the GC, $\psi$, is 
\begin{equation}
{\cal J}_{S}(\psi) = \frac{1}{r_{\odot} \rho_{\odot}^{2}} \int_{0}^{\ell_{max}}  \rho_{\chi}^{2}\left( r\left(\psi,\ell\right) \right) d\ell \, .
\end{equation} 
The integration limit is determined by the size of the MW DM halo: $\ell_{max} =  \sqrt{R^{2} - r_{\odot}^{2}\sin^{2}{\psi} }+r_{\odot} \cos{\psi} $, where $R = 200\, {\rm kpc}$ is the halo's virial radius. The J-factor is largely insensitive to the exact value of $R$. The galactocentric distance is $r(\psi,\ell) = \sqrt{r_{\odot}^{2}-2 \ell r_{\odot} \cos{\psi} +\ell^{2} }$. To compare the theoretical expectations with detector observables, one simply average the J-factor over the detector angular resolution or the field of view.  The J-factors for smooth halos are shown in Fig.~\ref{fig:plot_jfactor_LOW} and Fig.~\ref{fig:plot_jfactor_HIGH}.

In principle, the isotropic Galactic component of the IGRB from DM annihilation is given by the zeroth component of the spherical harmonic decomposition, or equivalently the average of the field of view of the observation. This might be complicated by the masking of the sky (e.g., the bright Galactic plane) and all the background reductions performed by the Fermi Collaboration. We therefore take the most conservative estimate by taking the constant J-factor from the Anti-GC ($\psi = \pi$), 
\begin{equation}
{\cal J}|_{\rm iso} \equiv \frac{1}{\int d\Omega}\int d\Omega\, {\cal J}_{S}(\pi) = {\cal J}_{S}(\pi)\,.
\end{equation}
A more detailed analysis for determining the isotropic Galactic component, possibly by including a DM template to the Fermi IGRB analysis, would further improve our DM constraint. 

\subsubsection{Galactic substructure enhancement}

In $\rm \Lambda CDM$, structures form hierarchically. The smallest DM halos are expected to form first. Some of these small halos subsequently merge and eventually may live in large host halos of galaxies or clusters. During structure formation, the small halos that are captured by larger halos are tidally disrupted and their low-density outer layers are stripped. The dense cores, however, could very well survive and become subhalos of the main halo \cite{Goerdt:2006hp,Berezinsky:2007qu,Afshordi:2009hn} (however, also see \cite{Angulo:2009hf}). We collectively define all of these surviving DM clumps to be substructures. High resolution simulations are beginning to resolve substructures down to the resolution limit \cite{Springel:2008cc, Diemand:2008in}. These substructures can lead to many interesting DM phenomenologies \cite{Koushiappas:2009du}.

\begin{figure}[t]
\includegraphics[angle=0.0, width=\columnwidth]{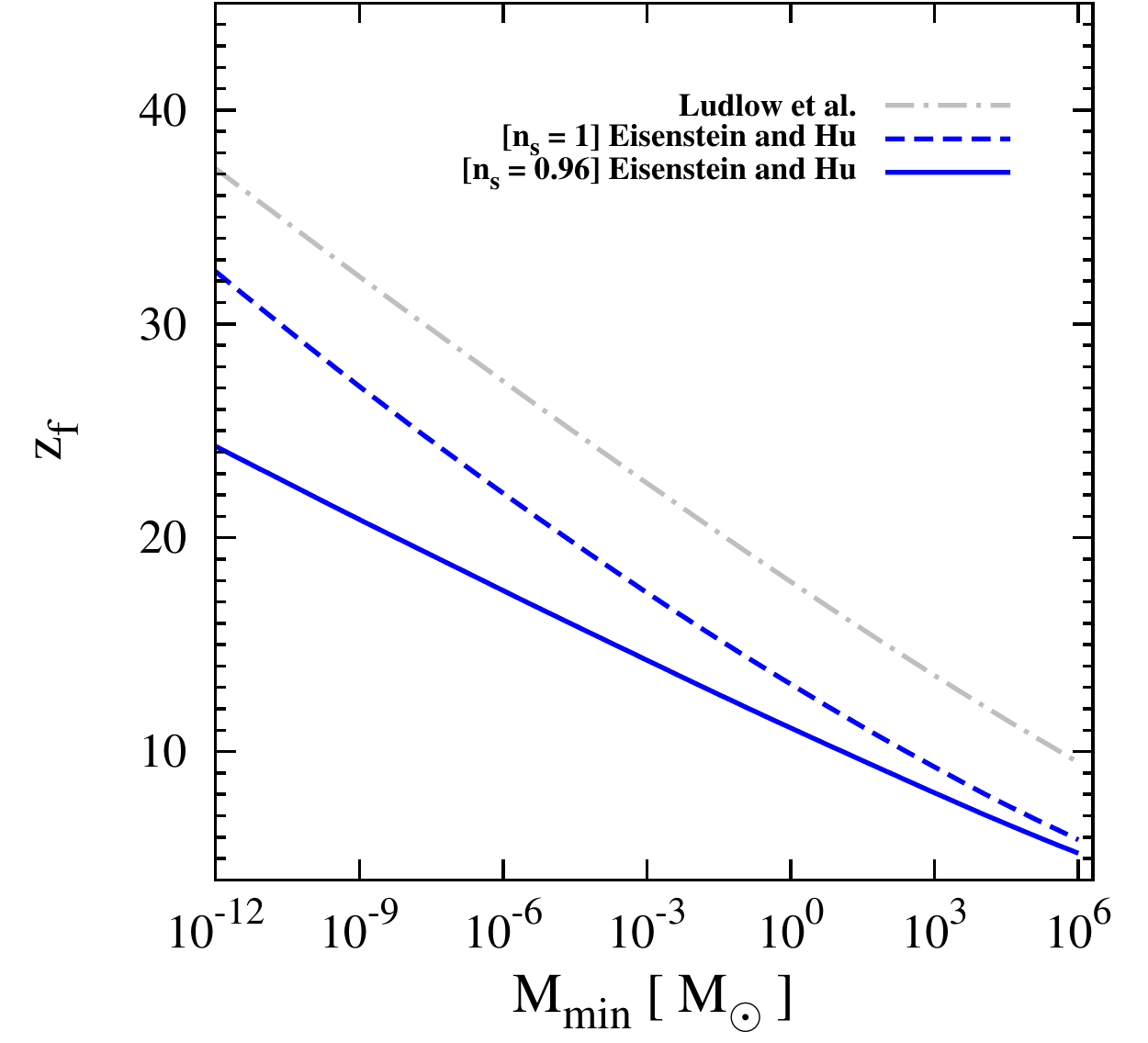}
 \caption{
The formation redshift, $z_{f}$, versus $ M_{\rm min}$, for the first generation halos. The formation redshift is obtained by requiring the linear mass variance be equal to the characteristic overdensity, $\sigma_{L}\left({ M},z \right) = 1.686$ \cite{Green:2005fa, Koushiappas:2009du}. For $\sigma_{L}\left({ M},z \right)$, we use the fitting functions of Eisenstein and Hu \cite{Eisenstein:1997jh}, which are evaluated and normalized with the Planck cosmological parameters \cite{Ade:2013zuv}. For illustration, we also show $z_{f}$ for the $n_{s}=1$ case as well as the extrapolated results from simulations by Ludlow et al.~\cite{Ludlow:2013bd} (we take $z_{f}$ to be $z_{-2}$ ).
} 
\label{fig:plot_formationz}
\end{figure}

\begin{figure}[t]
\includegraphics[angle=0.0, width=\columnwidth]{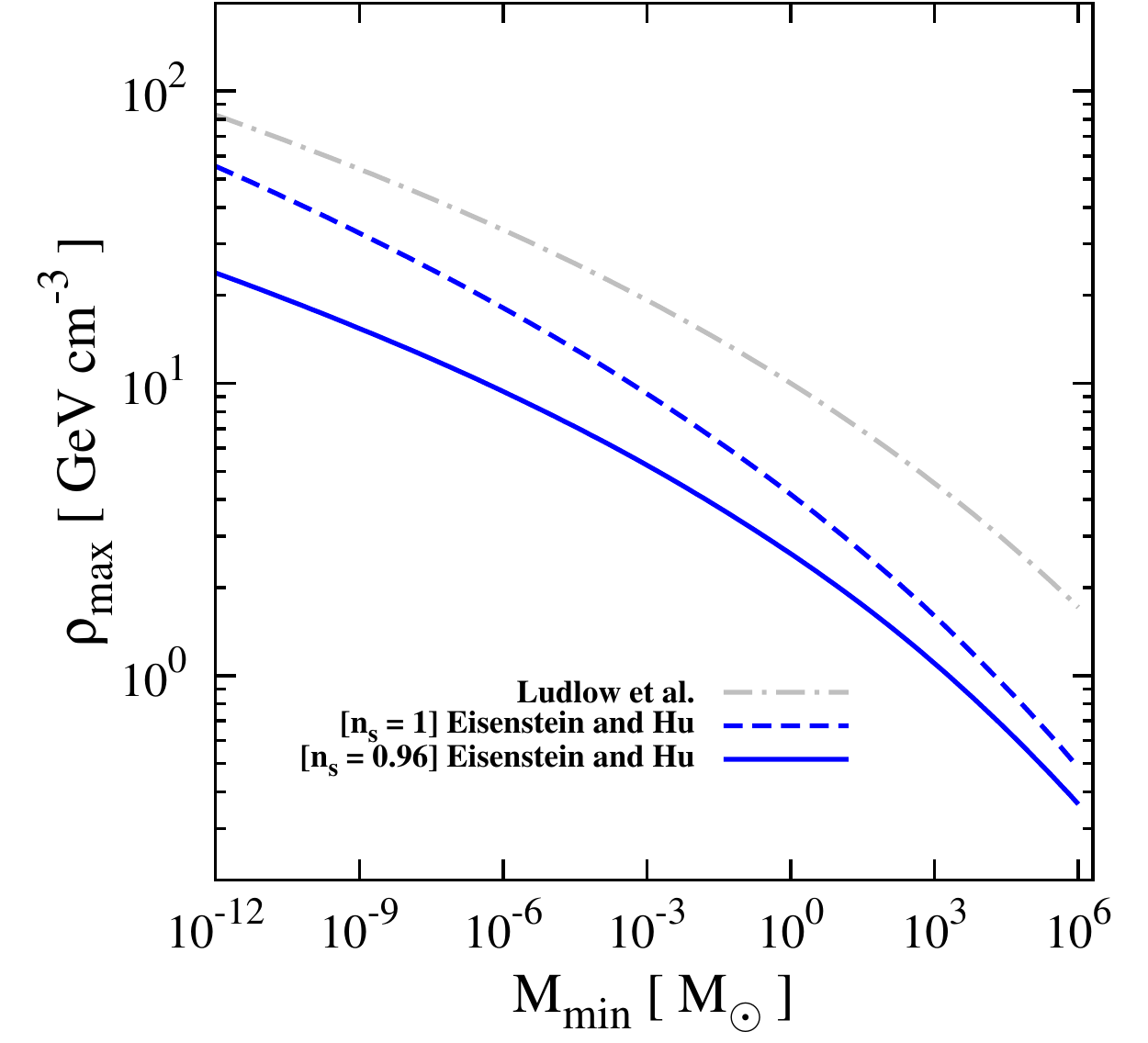}
\caption{
The characteristic density of the first generation halos, $\rho_{\rm max}$, versus $ M_{\rm min}$, for the corresponding cases of $z_{f}$ in Fig.~\ref{fig:plot_formationz} using Eq.~\ref{eq:rhomax}. The substructure boost is approximately proportional to $\rho_{\rm max}$ (Eq.~\ref{eq:galacticboost}). 
 }
\label{fig:plot_rhomax}
\end{figure}

Smaller DM structures tend to have higher concentrations \cite{Bullock:1999he}, which can be understood by their earlier formation time at which the background density is higher. Therefore, although substructures may not occupy much of the total volume of a halo, they could significantly enhance the DM annihilation rate of a halo.

To describe the substructure boost to the isotropic Galactic component of the IGRB, we use the theoretical model proposed by Kamionkowski and Koushiappas \cite{Kamionkowski:2008vw}. This model was later calibrated to high-resolution simulations by Kamionkowski, Koushiappas, and Kuhlen (\cite{Kamionkowski:2010mi}, hereafter K10), and therefore can be used to calculate the boost of the Galactic annihilation rate relative to a smooth halo density profile for the MW. The Galactic local boost factor as a function of $r$, for a velocity independent $\sigma v$, is

\begin{eqnarray} \label{eq:galacticboost}
      B(r) &=& f_s(r) e^{\delta^2_{f}} \nonumber \\
      & & +
      \left[1-f_s\left(r\right) \right]\frac{1+\alpha}{1-\alpha} \left[
      \left(\frac{\rho_{\rm max}}{\rho_{\chi}(r)} \right)^{1-\alpha}
      -1 \right], 
\end{eqnarray}
where $f_{s}(r)$ is the fraction of the volume that would be occupied by the smooth halo component, and $\rho_{\rm max}$ is the highest DM substructure density. The DM substructure fraction, $1-f_{s}$ is  
\begin{equation}\label{eq:oneminusf}
1-f_{s}(r) = \kappa \left[ \frac{\rho_{\chi}(r)}{\rho_{\chi}({\rm 100\; kpc})} \right]^{-0.26}. 
\end{equation}
The first term in $B(r)$ is the boost from the smooth halo component, by taking into account random fluctuations in the density. The second term describes the boost due to substructures. The substructure fraction is normalized by the parameter $\kappa$. Following K10, these parameters are determined to be $\delta_{f} = 0.2$, $\alpha = 0$, and $\kappa = 0.007$.

It has been pointed out in Fornasa et al.~\cite{Fornasa:2012gu} that the original K10 model tends to give a conservative substructure enhancement compared to other studies \cite{Springel:2008by, Gao:2011rf, Pinzke:2011ek}, mainly due to different methodologies. This can be reconciled by increasing the substructure survival fraction parameter, $\kappa$, from $0.007$ to $0.15-0.2$. In subsequent discussions, we refer to $\kappa = 0.007$ as the LOW substructure case as a conservative estimate, and $\kappa = 0.18$ as the HIGH substructure case as an optimistic scenario.

The boost factor is approximately proportional to the characteristic density of the minimum halo mass, $\rho_{\rm max}$. It depends on the natal concentration, $c_{0}$, the formation redshift, $z_f$, and the mass, ${ M_{\rm min}}$, of the first generation halos \cite{Kamionkowski:2010mi},  
\begin{equation} \label{eq:rhomax}
\rho_{\rm max}({ M_{\rm min}}) = \frac{1}{12} \frac{c_{0}^3}{{\rm ln}(1+c_0) -\frac{c_0}{1+c_0}} \Delta\, \rho_{c}\left(z_{f}\left({ M_{\rm min}}\right)  \right) \; ,
\end{equation}
where $\Delta = 200$ is the halo over-density. $\rho_{c}(z) = \rho_{c}(0) \mathcal{H}^{2}(z)$, where $\rho_{c}(0) = 1.05\times 10^{-5} h^{2} \, {\rm GeV\,cm^{-3} }$ is the critical density, and $\mathcal{H}^{2}(z) = \Omega_{\Lambda} + \Omega_{M}(1+z)^{3}$ is the Hubble function squared. The dependence of $ M_{\rm min}$ in $\rho_{\rm max}$ is mainly on $z_{f}({ M_{\rm min}})$, as the natal concentration is fairly constant for $z_{f} > 5$ \cite{Zhao:2008wd}. We follow K10 and take $c_{0}$ to be 3.5.

Parameter scans of some supersymmetric models show that $10^{-12}\,{ M_{\odot}}<{ M_{\rm min}}<10^{-3}\,{ M_{\odot}} $ \cite{Profumo:2006bv, Bringmann:2009vf}. Different models can drastically change the prediction for the value of ${ M_{\rm min}}$. We consider $10^{-12}$ and $10^{0}\, { M_{\odot}}$ as the lower and upper extreme cases for CDM, and adopt $10^{-6}\, { M_{\odot}}$  as the reference value. We consider $10^{6}\,{ M_{\odot}}$ unlikely for simple Cold DM models. Such a high value would require special DM physics (e.g. see Ref.~\cite{Aarssen:2012fx}, and also \cite{Laha:2013xua, Ahlgren:2013wba}) and is within the sensitivity of gravitational lensing probes \cite{Dalal:2001fq,2009MNRAS.398.1235X, 2010MNRAS.408.1969V}.

To estimate the value of ${\rm \rho_{max}}(M_{\rm min})$, we need to know the corresponding $z_{f}$. For the first generation halos, this can be estimated using cosmological perturbation theory \cite{Green:2005fa, Koushiappas:2009du}, since they are the first nonlinear structures of the Universe. Then $z_{f}$ is implicitly defined by $\sigma_{L}(M,z_{f}) = 1.686$, where 1.686 is the characteristic over-density of the 1-$\sigma$ linearized density fluctuation when halo collapse has occurred (see Ref.~\cite{SanchezConde:2006wu} and reference therein). $\sigma_{L}(M,z)$ is the linear mass variance defined by
\begin{equation}\label{eq:LinearMassVariance}
\sigma^{2}_{L}(M,z) = \int_{0}^{\infty} W^{2}(kR) \Delta^{2}_{L}(k,z) \frac{dk}{k} \;,
\end{equation}
where $W(kR)$ is the Fourier transform of the top hat window function, $\Delta^{2}_{L}$ is the dimensionless linear power spectrum, and $R$ is the comoving length scale. The mass of the collapsed halos can be estimated by $M \simeq \left(4/3\right)\pi R^{3}\rho_{c}(z_{f})$. We evaluate and normalize the mass variance using the fitting formula by Eisenstein and Hu \cite{Eisenstein:1997jh}, according to the Planck cosmological parameters \cite{Ade:2013zuv}. We also take into account the non-unity of the spectral index ($n_{s}=0.96$, without running), which is measured by the Planck collaboration with high significance. The effect of the slight tilt is amplified at small scales that we are interested in. Varying the index by approximately the 1-$\sigma$ Planck limit ($n_{s}=0.96\pm0.01$) yields a 5\% change in $z_{f}$ for ${ M_{\rm min}}=10^{-6}\,{ M_{\odot}}$, which translates into a 15\% change for $\rho_{\rm max}$.   We have considered only the 1-$\sigma$ density fluctuations which collapse into halos. Higher-$\sigma$ density fluctuations will collapse even earlier, and are thus denser, but they are correspondingly rarer.

In Fig.~\ref{fig:plot_formationz}, we show $z_{f}$ as a function of $ M_{\rm min}$ for  $n_{s}=0.96$. For comparison, we also show the case for $n_{s}=1$ and an extrapolation from the simulation of Ludlow et al.~\cite{Ludlow:2013bd}. The hierarchical nature of structure formation is apparent in this plot, with the smaller halos forming earlier. In Fig.~\ref{fig:plot_rhomax}, we show the corresponding $\rho_{\rm max}$ evaluated using Eq.~\ref{eq:rhomax}.


To incorporate the effect of substructure, we insert the boost factor into the line of sight integral to obtain the J-factor with substructure enhancement, ${\cal J}_{B}(\psi)$, 
\begin{equation}
{\cal J}_{B}(\psi) = \frac{1}{r_{\odot} \rho_{\odot}^{2}} \int_{0}^{\ell_{max}}  \rho_{\chi}^{2}\left[ r(\psi,\ell) \right] \cdot B\left[r(\psi,\ell) \right] d\ell \, .
\end{equation}
In Fig.~\ref{fig:plot_jfactor_LOW} and Fig.~\ref{fig:plot_jfactor_HIGH}, we show the effect of substructure on the J-factor for the LOW and HIGH substructure boost cases, respectively.

It is well known that the J-factor near the GC is very profile dependent \cite{Yuksel:2007ac}. However, substructures have relatively small enhancements to the J-factor at the GC, since substructures are more susceptible to tidal effects in high density regions. Therefore, DM signals from the GC can be considered to be substructure independent. The K10 substructure model qualitatively reflects this feature. However, the calibration to simulation inevitably breaks down near the GC, due to finite resolution effects \cite{Kamionkowski:2010mi}. Since details at the GC have no effect to our result, we assume the K10 model is valid at all regions.

As a result, any $\sigma v$ extracted from GC analysis is subjected to profile dependence, but independent of the underlying substructure assumptions. On the other hand, the J-factor is practically profile independent at large angles. We therefore find the isotropic Galactic component depends mostly on the substructure content of the halo, but not the density profile. The substructure enhancement for the isotropic Galactic component depends sensitively on the survival fraction $\kappa$. For the LOW (HIGH) substructure case, the boost is at most a factor of 1.5 (10).

It is also interesting to see that at $\sim 30^{\circ}$, the DM signal is the least uncertain relative to both density profile \cite{Yuksel:2007ac,Springel:2008by} and substructure scenarios. Therefore, one would ideally prefer to detect Galactic DM annihilation from such angles to minimize the astrophysical uncertainty on DM density distribution. 

\subsection{Extragalactic component}

The gamma-ray (number flux) intensity from extragalactic DM self-annihilation, $I_{\gamma}^{EG}(E_{0})$, is given by the cosmological line of sight integral, 
\begin{eqnarray} \label{eq1}
 &&I_{\gamma}^{\rm EG}(E_{0})  \\ 
\nonumber &=& \frac{{\sigma v} }{8 \pi}  \int \frac{v_c dz}{H_0 \mathcal{H}(z)} \frac{\langle\delta^{2}(z) \rangle}{(1+z)^{3}} \left(\frac{\bar{\rho}_{\chi}(z)}{m_{\chi}}\right)^{2} \frac{dN_\gamma}{dE}(E) e^{-\tau(z,E_0)} \\
\nonumber &=& 1.9\times 10^{-15}\, {\rm cm^{-2}\,s^{-1}\,sr^{-1}\,GeV^{-1}} \times \\
\nonumber && \left[ \frac{\sigma v}{2.2\times 10^{-26} \, {\rm cm^{3}\,s^{-1}} } \right] \left[ \frac{100\,{\rm GeV}}{m_{\chi}} \right]^{2} \times \\
\nonumber &&  \int{dz \frac{(1+z)^3}{\mathcal{H}(z)} } \langle\delta^{2}(z) \rangle \frac{dN_\gamma}{dE/{\rm GeV}}(E) e^{-\tau(z,E_0)}\; ,
\end{eqnarray}
where $E$ is the center-of-momentum frame energy given by $E = E_{0}(1+z)$, $v_{c}$ is the speed of light, and $\bar{\rho}_{\chi}(z)$ is the cosmological mean DM density. The clumping factor, $\langle\delta^{2}(z) \rangle$, which measures the cosmologically averaged DM density squared, relative to the mean DM density squared, $\langle\delta^{2}(z) \rangle = \langle\rho_{\chi}^{2}(z) \rangle / \bar{\rho}_{\chi}^{2}(z)$.
High-energy gamma rays propagating through intergalactic space will suffer attenuation due to the Extragalactic Background Light (EBL). This effect is included in the attenuation factor, $e^{-\tau(E_{0},z)}$ (see Sec.~\ref{sec:ebl}). 

The clumping factor is the main theoretical astrophysical uncertainty in evaluating the expected DM annihilation intensity. We review how to evaluate the clumping factor using the Halo Model approach, with or without substructures in massive halos. We also review how to evaluate the equivalent quantity using the Power Spectrum approach, which is complementary to the halo model approach in terms of theoretical uncertainties.

\begin{figure}[t]
\includegraphics[angle=0.0, width=\columnwidth]{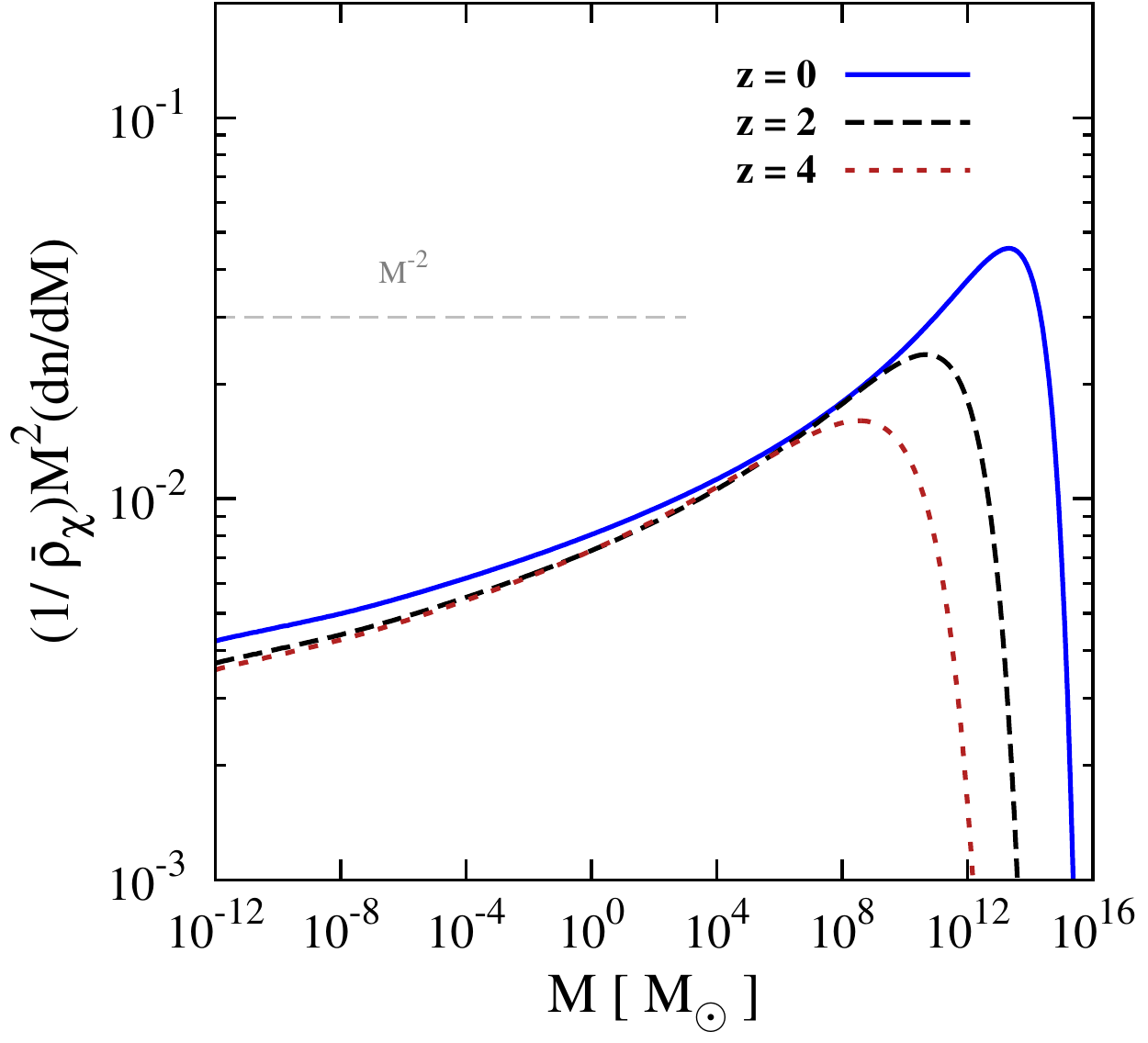}
\caption{The normalized halo mass function $(1/{\bar{\rho}_{\chi} })M^{2}dn/dM $ plotted versus $ M$ for redshift $z=0, 2, 4$. The halo mass function as a function of the linear mass variance is given by P12 \cite{Prada:2011jf}. The redshift evolutions of the fitting parameters are given by Tinker et al.~\cite{Tinker:2008ff}.} 
\label{fig:plot_halomassfunction}
\end{figure}

\begin{figure}[t]
\includegraphics[angle=0.0, width=\columnwidth]{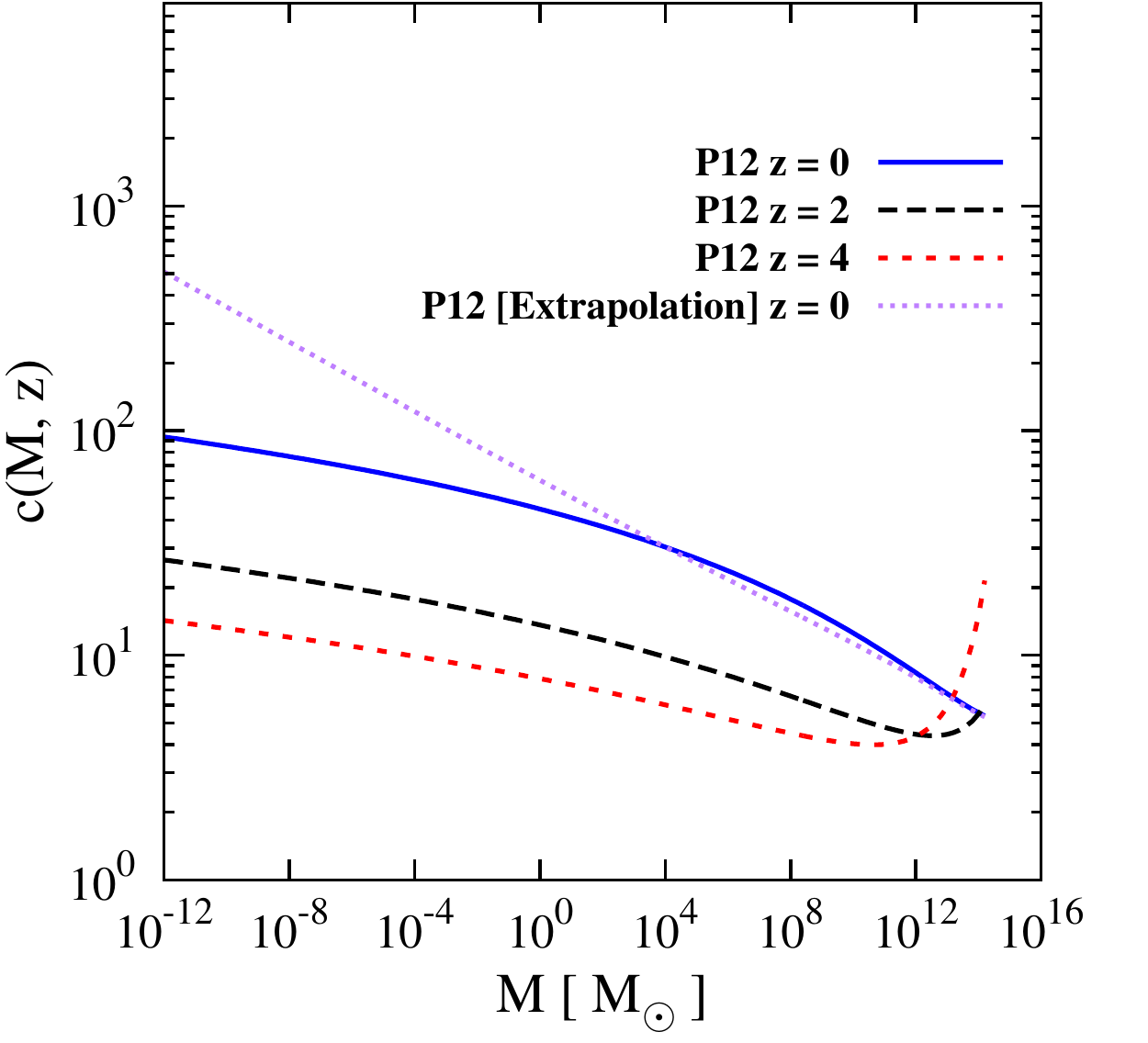}
\caption{The concentration parameter, $c(M,z)$, plotted against $M$ for redshift $z=0, 2, 4$. The concentration mass relation as a function of the linear mass variance is again given by P12 \cite{Prada:2011jf}. For comparison, we also show the concentration if we simply extend the concentration-mass relation to small scales using the analytic function given in P12. 
}
\label{fig:plot_concentration}
\end{figure}


\subsubsection{Halo Model approach with smooth halos only}
The clumping factor for smooth halos, $\langle\delta^{2}_{S}(z) \rangle$, can be calculated using the Halo Model framework \cite{Bergstrom:2001jj, Ullio:2002pj},  
\begin{eqnarray}\label{eq:clumpingfactor}
 \langle\delta_{S}^{2}(z) \rangle &=& \frac{\langle\rho_{\chi}^{2}(z) \rangle}{\bar{\rho}_{\chi}^{2}(z)} \\ 
\nonumber&=& \frac{1}{\bar{\rho}_{\chi}^{2}(z)} \int{dM} \frac{dn}{dM}(z) \int_{r<R}{dV} \rho^{2}_{\chi}(\rho_{s},r_{s})
\end{eqnarray}
where $ dn/dM$ is the distinct halo mass function in physical units, which describes the number density of bound objects with mass $M$ at a particular redshift. These objects are distinct halos --- their centers are not inside the virial radius of larger halos. The density profile of a particular halo, $\rho_{\chi}(\rho_{s},r_{s})$, is characterized by its scale density, $\rho_{s}$, and scale radius, $r_{s}$, which in turn depends on the halo mass, $ M$, and redshift, $ z$.  In Eq.~\ref{eq:clumpingfactor}, $\langle {\rho^{2}} \rangle$ denotes the volume average of halos of all masses of the density squared and $\bar{\rho}$ denotes the cosmic mean density. 

We use the distinct halo mass function from Prada et al.~(\cite{Prada:2011jf}, hereafter P12). The P12 halo mass function is obtained by fitting to four cosmological simulations. The fitting functional form follows from the Press-Schechter theory and its extensions \cite{Press:1973iz, Sheth:1999mn}. The halo mass function describes the full hierarchy of distinct cosmological DM halos down to $ M_{\rm min}$, and the cosmology dependence enters through the linear mass variance, $\sigma_{L}$ (Eq.~\ref{eq:LinearMassVariance}). For the redshift dependence of the fitting parameters, we follow those from Tinker et al.~\cite{Tinker:2008ff}. In Fig.~\ref{fig:plot_halomassfunction} we show the halo mass function for several redshifts. The low mass dependence is slightly harder than the critical $ M^{-2}$ behavior. 

The volume integral of the density squared can be simplified using two halo mass relations, which convert the $\{\rho_{s}, r_{s}\}$ dependence to only the concentration parameter, $c(M,z) = R/{r_{s}}$. The first one is 
\begin{equation}
M = \frac{4}{3}\pi R^{3} \Delta \rho_{c}(z) \;,
\end{equation}
where $ R$ is the virial radius of the halo. The second halo mass relation is 
\begin{eqnarray}
M &=& \int_{r<R}{dV}\rho_{\chi}(\rho_{s},r_{s})
\end{eqnarray}
which is integrated up to the virial radius. 

The clumping factor can now be written as 
\begin{eqnarray}\label{equation:clumping1}
 \nonumber \langle\delta_{S}^{2}(z,{ M_{\rm min}}) \rangle  
&=& \frac{1}{\Omega_{\chi}} \int_{ M_{\rm min}} dM  \frac{\mathcal{H}(z)^{2}}{(1+z)^{3}}\frac{1}{\bar{\rho}_{\chi}(z)} \frac{dn}{dM} \frac{M\Delta}{3}\\
&& \times \int{d\hat{c}}\; P(c,\hat{c}) \hat{c}^{3} \frac{I_{2}(\hat{c})}{I_{1}(\hat{c})^{2}}\; ,
\end{eqnarray}
where we have introduced the dimensionless integral $I_{n}(c) = \int_{0}^{R}(dr/r_{s})\,(r/r_{s})^{2}({\rho_{\chi}(r)}/{\rho_{s}})^{n} $, and the log-normal distribution, $P(c,\hat{c})$, with constant 1-$\sigma$ deviation $\sigma_{log_{10}}=0.13$ (or $\sigma_{ln}=0.3$) \cite{Sheth:2004vb,Neto:2007vq} around the mean concentration parameter, $c(M,z)$.  We simplify the formalism by defining the effective cut-off in the Halo Mass function to be the minimum halo mass, $ M_{\rm min}$, thus ignoring objects with masses below $ M_{\rm min}$ \cite{Bertschinger:2006nq}. 

We argue that this definition of $ M_{\rm min}$ is effectively equivalent to the $ M_{\rm min}$ in the Galactic substructure calculation. The smallest substructure mass in halos may be less than the smallest cosmological halo mass because of tidal disruption in merging. But the relevant part of DM annihilation, which is the maximum density of the substructures, can be assumed to be unaffected by tidal disruptions \cite{Goerdt:2006hp,Berezinsky:2007qu,Afshordi:2009hn}. 

The last ingredient we need is the mean concentration parameter $c(M,z)$, which is a quantitative measure of halo concentrations. We use the analytic function from P12, which is derived from cosmological simulations and agrees well with cluster observations. The P12 result for $c(M,z)$ shows a remarkably tight relation with the linear matter mass variance, $\sigma_{L}(M)$, for which we again use the linear mass variance given by Eisenstein and Hu with the Planck cosmology. It is intuitive that halo concentrations would tightly correlate with the linear mass variance, since the latter is intimately related to halo formation \cite{Bullock:1999he}. 

We show the concentration-mass relation for NFW profiles in Fig.~\ref{fig:plot_concentration} for $z=0, 2, 4$. For comparison, we also show the concentration if we simply extend the fitting function for $\sigma_{L}(M)$ from P12 to small scales. Recent microhalo simulations have shown that a $\sim 10^{-7}\, { M_{\odot}}$ first generation halo has concentration $ 57<c<84 $ at redshift zero \cite{Anderhalden:2013wd}, with a mean value of 72. We find that a naive substitution of the linear mass variance from \cite{Eisenstein:1997jh} slightly underestimates the concentration at small mass scales. Therefore, we change one of the fitting parameter ($c$ in Eq.~(16) in \cite{Prada:2011jf}) from 1.022 to 1.05. The resulting concentration increases from 67 to 73 at $10^{-7}\, { M_{\odot}}$, with negligible changes at large scales. From Fig.~\ref{fig:plot_concentration}, we can see the concentrations grow slower than a naive power-law extrapolation to small halo masses. This ``bending'' of concentration is also recently noted by Ludlow et al. \cite{Ludlow:2013vxa} and by S\'anchez-Conde and Prada \cite{Sanchez-Conde:2013yxa}.

The rising concentration at large mass scales is a novel feature from the simulation \cite{Klypin:2010qw}. This feature, though interesting, has no effect on our calculation due to the rapidly falling halo mass function at the corresponding mass and redshift.

\begin{figure}[t]
\includegraphics[angle=0.0, width=\columnwidth]{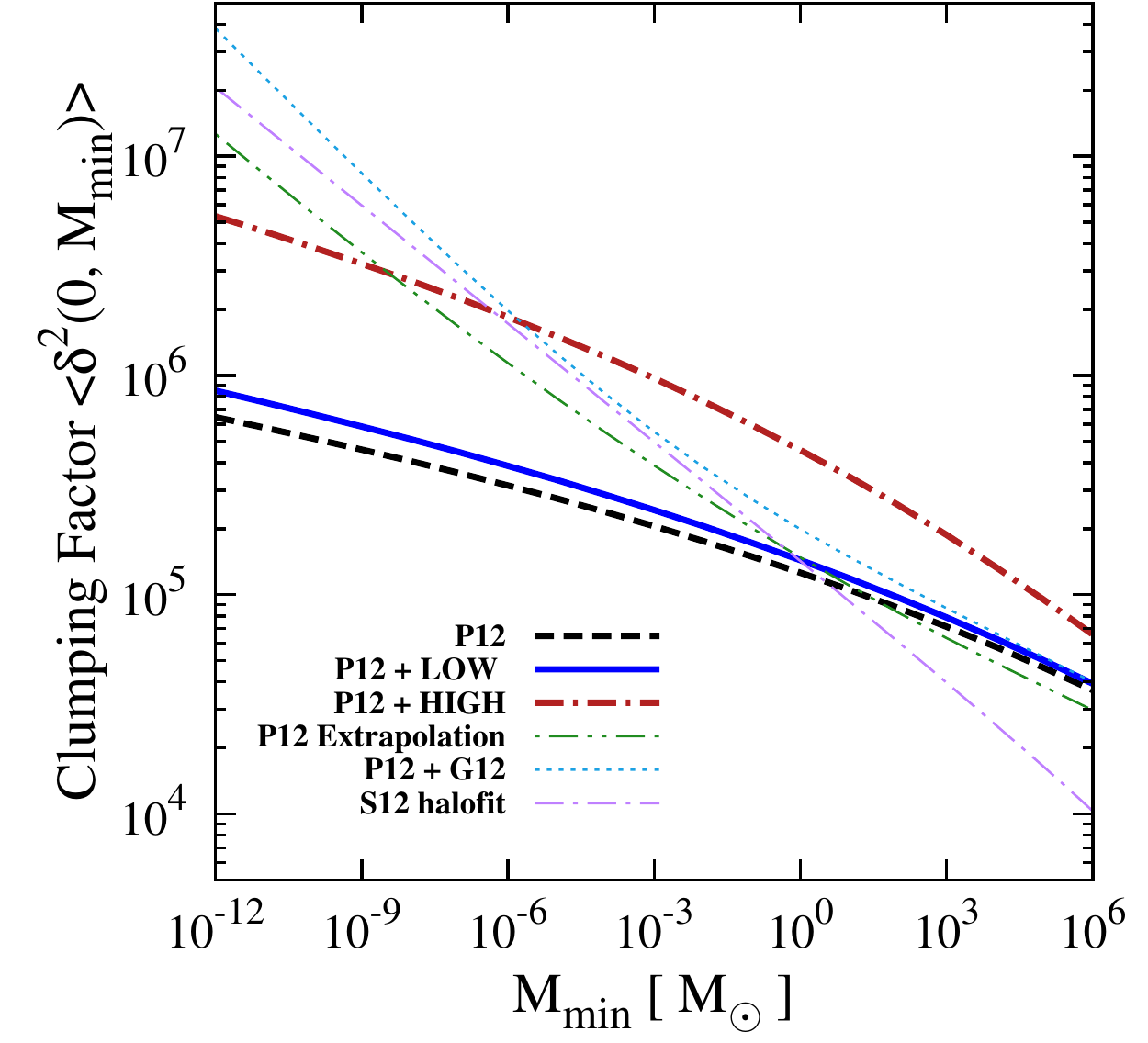}
\caption{
The clumping factor at $z=0$ versus $ M_{\rm min}$.  Using the P12 \cite{Prada:2011jf} Halo Model, we obtain the clumping factor without substructure enhancement. Adding the K10 \cite{Kamionkowski:2010mi} substructure model, we show the substructure-enhanced clumping factor for LOW and HIGH cases. For comparison, we also show the clumping factor if we simply extrapolate the concentration relation in P12, the clumping factor with G12 \cite{Gao:2011rf} substructure model, and the clumping factor using the extrapolated $halofit$ \cite{Smith:2002dz} non-linear power spectrum, following S12 \cite{Serpico:2011in}. In this work, we consider the LOW and HIGH scenarios as the conservative and optimistic substructure cases.  
  } 
\label{fig:plot_clumpingfactor}
\end{figure}

In Fig.~\ref{fig:plot_clumpingfactor}, we show the clumping factor as a function of $ M_{\rm min}$, using the P12 model with $\sigma_{L}(M)$ from \cite{Eisenstein:1997jh}. At the extreme case of $ M_{\rm min} = 10^{6}\, M_{\odot}$, the P12 model yields $\sim 4\times 10^{4}$, consistent with similar evaluations \cite{Ando:2005hr,Beacom:2006tt, Zavala:2009zr}. This mass scale is within the simulation limits \cite{Zavala:2009zr} , and is also within the reach of gravitational lensing probes \cite{Dalal:2001fq,2009MNRAS.398.1235X, 2010MNRAS.408.1969V}. Therefore, we consider this to be the minimum DM clustering value, a lower bound to the clumping factor.  

Microhalo simulations show that first generation halos have a steeper inner slope than the normal NFW profiles \cite{Ishiyama:2010es,Anderhalden:2013wd}. This would enhance the annihilation signal from microhalos. One may also be interested in the profile dependence of the clumping factor. It is not straightforward, however, to change the density profile in calculating the clumping factor, since the value of the concentrations extracted from simulations depend on the assumed profile \cite{Prada:2011jf}. Nonetheless, the clumping factor is expected to be relatively insensitive to the density profile. For example, the total annihilation luminosity from the MW halo only experience a change of -20\% or +30\%, if isothermal or Einasto profiles are used. For simplicity, we use the NFW profile for all the evaluations.

\subsubsection{Halo Model approach with substructures}

In the above calculation, we assume each halo in the halo mass function has a universal smooth DM density profile. However, in addition to the cosmological isolated small halos, substructures within halos also contribute to the clumping factor. 

Unlike the Galactic case, the observer is outside of all the halos observed and each halo has different mass and size. To incorporate substructure effects, we extend the K10 substructures model to different halo sizes, following the approach taken by S\'anchez-Conde et al.~\cite{SanchezConde:2011ap} (also see Fornasa et al.~\cite{Fornasa:2012gu}). To recalibrate the K10 model for different halo sizes, the substructure fraction needs to be modified,  
\begin{equation}
1-f_{s}(r) = \kappa\left( \frac{\rho_{\chi}(r)}{\rho_{\chi}(3.56\times r_{s})} \right). 
\end{equation}
In doing so we have assumed the same radial dependence of $f_{s}(r)$ for all halo masses. The factor of 3.56 is the conversion factor from the Milky Way halo to the size of the simulation from K10. The local boost factor would enter inside the dimensionless integral $I_{2}$ in Eq.~\ref{equation:clumping1}. The substructure enhanced clumping factor, $\langle\delta_{B}^{2}(z,{ M_{\rm min}}) \rangle$, is
\begin{eqnarray}
&&\langle\delta_{B}^{2}(z,{ M_{\rm min}}) \rangle \\
\nonumber =&& \frac{1}{\Omega_{\chi}} \int_{ M_{\rm min}} dM  \frac{\mathcal{H}(z)^{2}}{(1+z)^{3}}\frac{1}{\bar{\rho}_{\chi}(z)} \frac{dn}{dM} \frac{M\Delta}{3}\int d\hat{c}\; P(c,\hat{c}) \hat{c}^{3} \frac{\tilde{I}_{2}}{I_{1}^{2}}\; ,
\end{eqnarray}
where
\begin{equation}
\tilde{I}_{2}(c,{ M_{\rm min}}) = \int_{0}^{c}d\frac{r}{r_{s}} \left(\frac{r}{r_{s}} \right)^{2} \left(\frac{\rho_{\chi}(r)}{\rho_{s}} \right)^{2}\cdot B(r, { M_{\rm min}}) .
\end{equation}
Recall that the $ M_{\rm min}$ dependence enters $ B(r, { M_{\rm min}})$ through $ \rho_{max}$. 

In Fig.~\ref{fig:plot_clumpingfactor}, we show the clumping factor with substructures, $\langle\delta_{B}^{2}(0,{ M_{\rm min}}) \rangle$, for the LOW and HIGH substructure cases. For the LOW case, the substructure boost is small. For the HIGH case, the substructure boost is more important, where the enhancement ranges from about a factor of 2 in large $ M_{\rm min}$ to a factor of 6 in the smallest $ M_{\rm min}$. 

For extragalactic halos one can also use the substructure boost by Gao et al.~(\cite{Gao:2011rf}, hereafter G12), which corresponds to the HIGH case in cluster scale. The substructure fraction for HIGH is tuned to match the G12 boost factor for an individual cluster scale halo \cite{Han:2012au,Dasgupta:2012bd,Murase:2012rd}, with $ M_{\rm min}$ assumed to be $10^{-6} \, { M_{\odot}}$. Taking into account the scaling factor of the G12 boost factor to $ M_{\rm min}$ ($\propto { M_{\rm min}^{-0.226}}$ \cite{Springel:2008by, Han:2012au}), the clumping factor for G12 is shown in Fig.~\ref{fig:plot_clumpingfactor}. The clumping factor is even higher than the HIGH case for $ M_{\rm min} $ less than about $10^{-6} \, { M_{\odot}}$, but opposite otherwise. We see that even for the HIGH substructure case, the clumping factor is more conservative than the the cases where power-law extrapolation is used, due to a slower increase in small-scales. 

The shape of the clumping factor is inherited from both the shape the halo mass function and the concentration-mass relation. The characteristic shape of the P12 clumping factors are due to the slower increase in concentration in smaller masses (Fig.~\ref{fig:plot_concentration}), which ultimately traces back to the flattening of $\Delta^{2}_{L}$ in $\sigma_{L}(M)$ (Eq.~\ref{eq:LinearMassVariance}). Physically, this reflects the property that small halos over a large range of mass formed in a relatively small period of time. As a result, when predicting the clumping factor (i.e., the extragalactic DM annihilation flux), decreasing $M_{\rm min}$ leads to only a small increase in the clumping factor. In contrast, when constraining $M_{\rm min}$, a small improvement on the flux limit would lead to a large improvement on the limit for $M_{\rm min}$. 

Additional clustering of DM can also be achieved by density spikes near Black Holes \cite{Gondolo:1999ef,Horiuchi:2006de} or adiabatic contraction of DM halos \cite{Gnedin:2004cx}. On the other hand supernova feedback might introduce a core to the density profiles for larger size halos \cite{Zolotov:2012xd,Brooks:2012vi,GarrisonKimmel:2013aq}. Warm or mixed DM, DM interactions with themselves \cite{Spergel:1999mh,Rocha:2012jg, Peter:2012jh}, or with other particles \cite{Aarssen:2012fx} can also significantly change dark matter density distributions. These effects deserve detailed studies and are outside the scope of this work.

\subsubsection{Power Spectrum approach}

We have shown how to obtain the clumping factor using the Halo Model formalism. It can alternatively be obtained from the $r \to 0$ limit of the two-point correlation function, $\langle \delta(x+r)\delta(x) \rangle$, as shown by Serpico et al.~(\cite{Serpico:2011in}, hereafter S12). It can be expressed as an  integral of the non-linear power spectrum, $\Delta_{\rm NL}(k,z)$, 
\begin{equation}
\langle \delta^{2}(z) \rangle = \lim_{r\to 0}\int_{0}^{\rm k_{max}} \frac{dk}{k} \frac{\sin{kr}}{kr} \Delta_{\rm NL}(k,z)\;,
\end{equation}
where $\rm k_{max}$ is the cut-off of the non-linear power spectrum and corresponds to $  M_{\rm min}$ by ${ M_{\rm min}}\approx(4/3)\pi (\pi/ {\rm k_{max}})^{3} \bar{\rho}_{\chi}$.

Using the Power Spectrum approach, one has the obvious advantage that many uncertainties of the Halo Model are collectively reflected in the non-linear part of the power spectrum. The constraints from DM observation can be related to constraints on the shape and cut-off of the non-linear power spectrum $\Delta_{\rm NL}(k,z)$. It is, however, difficult to probe the small-scale non-linear regime in theoretical treatments and simulations, and the physics in Fourier space is more difficult to translate to physics in real space. 

Nonetheless, the power spectrum approach is appealing for its simplicity and different systematics. We selected the $halofit$ model \cite{Smith:2002dz} following S12 and extrapolate it to the scales relevant for our discussion of $ M_{\rm min}$. The resulting clumping factor is shown in Fig.~\ref{fig:plot_clumpingfactor}. The Power Spectrum result roughly agrees with the Halo Model approach.

\begin{figure}[t]
\includegraphics[angle=0.0, width=\columnwidth]{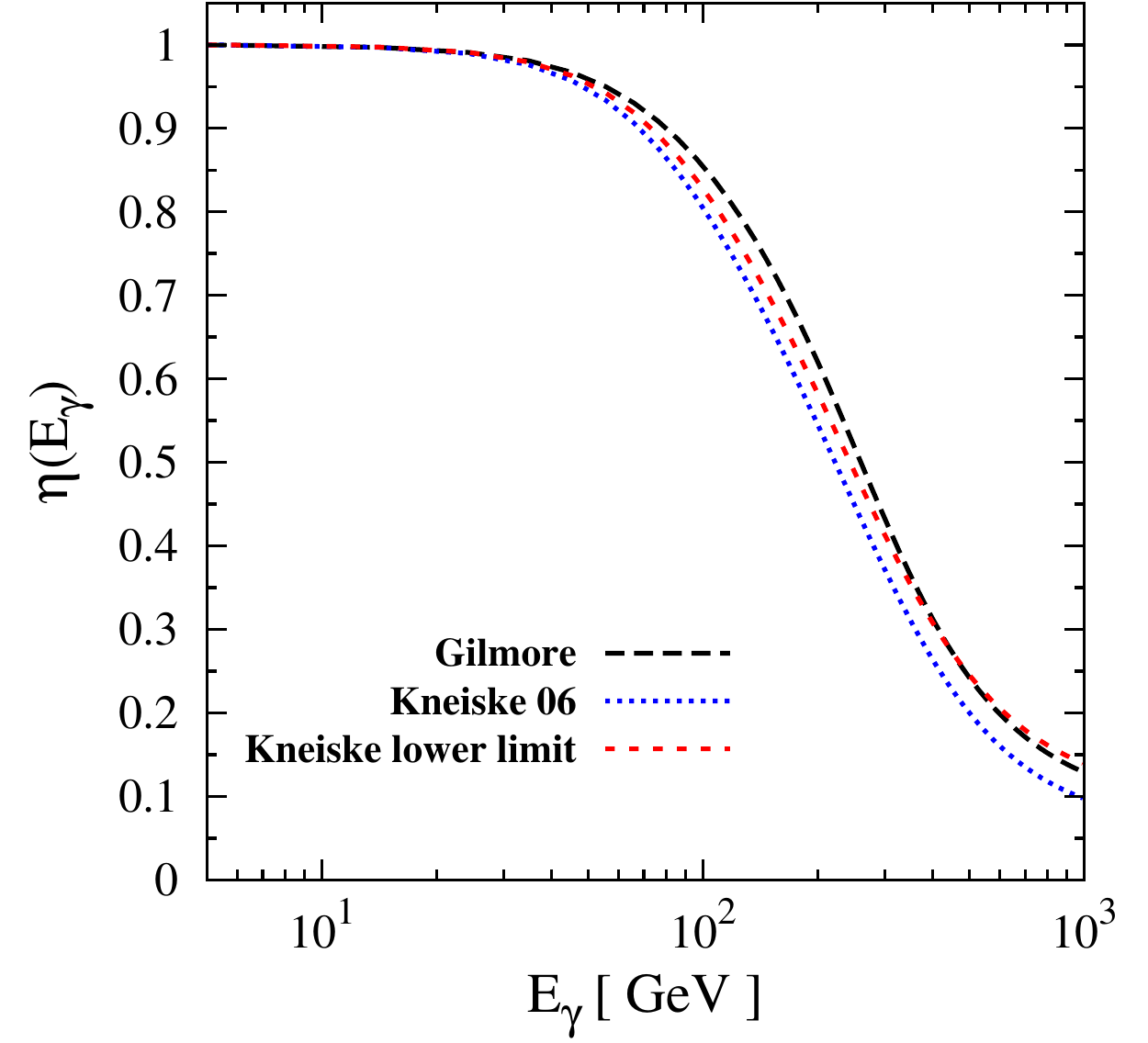}
\caption{The attenuation fraction of monochromatic gamma-ray signals from cosmological DM annihilation sources (Eq.~\ref{eq:eta}) versus the emitted photon energy for different EBL models.  We consider the EBL model from Gilmore et al.~\cite{Gilmore:2011ks}, ``Best Fit 06" from Kneiske et al.~\cite{Kneiske:2003tx}, and the ``Lower-Limit" from Kneiske and Dole \cite{Kneiske:2010pt}. All models considered are within 2-$\sigma$ of the Fermi measurement \cite{Ackermann:2012sza}. We adopt the Gilmore model throughout. } 
\label{fig:plot_attenuation}
\end{figure}

\begin{figure}[t]
\includegraphics[angle=0.0, width=\columnwidth]{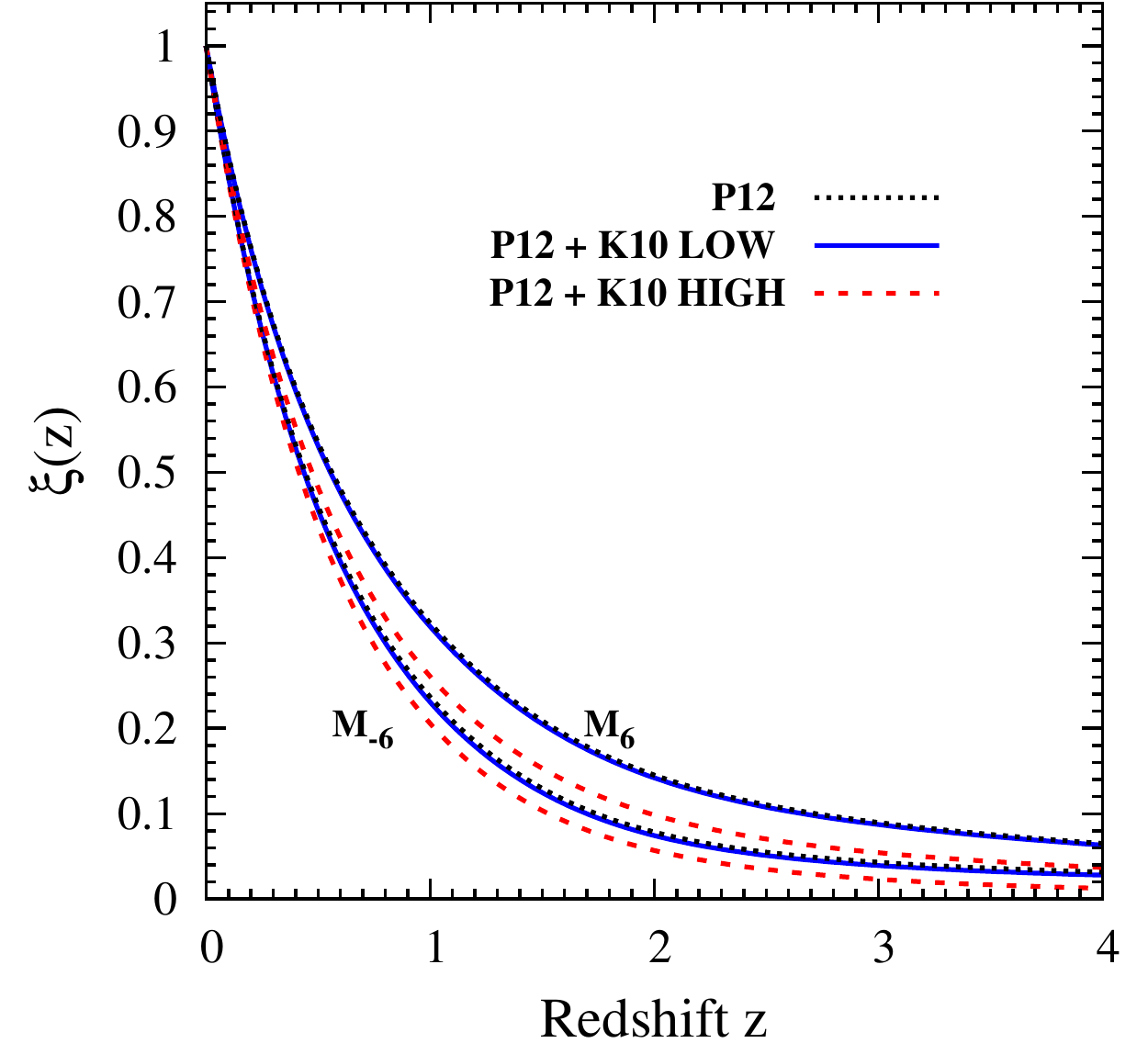}
\caption{The extragalactic DM annihilation redshift distribution (Eq.~\ref{eq:xi}). We show the distributions for the P12 Halo Model, P12 Halo Model with K10 LOW, and HIGH substructure models. The upper set of three lines uses ${ M_{\rm min}} = 10^{-6}\,{ M_{\odot}}$, while the the lower set uses ${ M_{\rm min}} = 10^{6}\,{ M_{\odot}}$. The shape of the distribution varies mildly in different scenarios. Most of the DM annihilation signal comes from small redshifts.   
} 
\label{fig:plot_profileshape}
\end{figure}

\subsection{EBL attenuation and redshift distribution}\label{sec:ebl}

In this section we discuss the effect of the EBL attenuation and the general redshift behavior of the extragalactic DM annihilation signal. 

\subsubsection{EBL attenuation}

It is evident from Eq.~\ref{eq1} that all the astrophysical and cosmological uncertainties are contained in the combination, $(1+z)^{3}(\langle\delta^{2}(z)\rangle/\mathcal{H}(z) )e^{-\tau(z,E_{0})}$. In particular, only $\tau(z,E_{0})$ depends on the nature and energy of the messenger, therefore also on the mass and the annihilation channel of DM particles. It is instructive to explore the effect of the EBL attenuation by looking at the ratio of the attenuated total flux to the unattenuated photon flux, $\eta(E_{\gamma})$, for monoenergetic photon emission,
\begin{eqnarray}\label{eq:eta}
&&\eta(E_{\gamma})  \\
\nonumber &&= \frac{\int{dE_{0}}\int{dz} \frac{(1+z)^{3}}{\mathcal{H}(z)}\langle \delta^{2}(z) \rangle \delta(E_{0}(1+z)-E_{\gamma})e^{-\tau(z,E_{0})}    }
{\int{dE_{0}}\int{dz} \frac{(1+z)^{3}}{\mathcal{H}(z)}\langle \delta^{2}(z) \rangle \delta(E_{0}(1+z)-E_{\gamma})} \; ,
\end{eqnarray}
where $\delta\left(E_{0}\left(1+z\right) - E_{\gamma}\right)$ is the Dirac-delta function connecting the observed energy and the emitted energy. This factor represents the relative flux suppression due to EBL absorption, according to DM clumping evolution. 

In Fig.~\ref{fig:plot_attenuation} we show $\eta(E_{\gamma})$ for a few different EBL models that are compatible with the latest Fermi results \cite{Ackermann:2012sza}. We use the P12 Halo Mass model with ${ M_{\rm min}} = 10^{-6}\,{ M_{\odot}}$ to evaluate $\eta(E_{\gamma})$. Different models share the same generic feature that attenuation affects annihilation signals with gamma-ray above $\sim50$ GeV, and the amount of attenuation is fairly consistent for different models. Throughout this work we adopt the Gilmore fixed model \cite{Gilmore:2011ks}, which has a slightly lower EBL compared to other models. The EBL mildly attenuates the 130 GeV DM, but has virtually no effect on light DM.

\subsubsection{Redshift distribution}

Another interesting quantity to see is the redshift distribution of the DM annihilation signal. We quantify this by defining the dimensionless quantity $\xi(z)$, 
\begin{eqnarray} \label{eq:xi}
\nonumber \xi(z) &=& \frac{ \frac{d}{dz}\int{dE_{0}}I^{\rm EG}_{\gamma}(E_{0})  }
{\frac{d}{dz}\int{dE_{0}}I^{\rm EG}_{\gamma}(E_{0})|_{z=0} } \\
&=& \frac{(1+z)^{3}}{\mathcal{H}(z)} \frac{\langle \delta^{2}(z) \rangle}{\langle \delta^{2}(0) \rangle } \frac{1}{1+z} \;.
\end{eqnarray}
Physically, $\xi(z)$ is the relative DM annihilation signal per redshift interval. To make the discussion independent of particle physics, we integrate out the energy spectrum, which results in the additional factor of $1/(1+z)$ at the end. This factors accounts for the energy binning effect or equivalently the cosmological time dilation, as the energy-time element is redshift invariant ($dt dE = dt_{0}dE_{0}$).  We also neglect the attenuation factor, $\tau$, to make the discussion independent of the EBL model and the annihilation products being observed.

We show $\xi(z)$ in Fig.~\ref{fig:plot_profileshape} for three cases: the P12 smooth Halo Model and, the P12 model with K10 substructure for LOW and HIGH cases. For a fixed $ M_{\rm min}$, substructure has minor effect on the distribution. Varying the value of $  M_{\rm min}$ also changes the shape slightly. In all cases, $\xi(z)$ is peaked at redshift zero. In terms of implications for detection prospects, not only does low-$z$ region have a larger flux, the less redshifted energy also means the signal is more detectable. This argument is even stronger if there is a considerable cosmic attenuation effect.  

The shape of $\xi(z)$ determines the gamma-ray profile of DM annihilation signals before detector smearing and cosmic attenuation. It encodes the redshift evolution of DM density distribution. Therefore, the signal profile with energy could be a probe for the cosmic structure evolution. The effect is however secondary to the signal strength, and we encourage future works to explore this possibility.

\subsection{Isotropic Galactic vs. Extragalactic}

One can compare the relative importance of the isotropic Galactic component to the extragalactic component (see also \cite{Yuksel:2007ac}), 
\begin{eqnarray}
\frac{I^{\rm EG}}{I^{\rm G}} \sim \left[\frac{\langle \delta^{2} \rangle}{8 \times 10^{4}}\right]\left[\frac{0.4}{\cal J}\right]   \,.
\end{eqnarray}
The J-factor for the isotropic Galactic component has a robust lower limit, ${\cal J} \sim 0.4$, for the case of no substructure at anti-GC. In this case, the extragalactic component will be comparable to the isotropic Galactic component, with $\langle \delta^{2} \rangle \sim 8\times 10^{4}$. This corresponds to $ M_{\rm min} \sim 10^{3}\, { M_{\odot}}$ for LOW substructure case, or $\sim 10^{5}\, { M_{\odot}}$ for HIGH substructure case. 

The isotropic Galactic component naturally breaks the ``clumping factor--$\sigma v$'' degeneracy. This can be seen by the following schematic equation, 
\begin{equation}\label{eq:schematic}
\Phi^{\rm IGRB} \propto \sigma v \left( \langle \delta^{2} \rangle \Phi_{0}^{\rm EG} + {\cal{J}} \Phi_{0}^{\rm G} \right)\,,
\end{equation}
where $\Phi^{\rm IGRB}$ is the total DM annihilation contribution to IGRB and $\Phi^{\rm EG}_{0}$ ($\Phi^{\rm G}_{0}$) is the extragalactic (isotropic Galactic) component properly normalized to factor out the dependence of $\sigma v$, $\langle \delta^{2} \rangle$, and $\cal J$.  For large $\langle \delta^{2} \rangle$, or small $ M_{\rm min}$, the isotropic Galactic component is negligible and the $\langle \delta^{2} \rangle \sigma v$ degeneracy is apparent. For small $\langle \delta^{2} \rangle$, or large $ M_{\rm min}$, the isotropic Galactic component dominates. In particular for large $ M_{\rm min}$, the substructure enhancement to the J-factor is small, thus the degeneracy is naturally broken. In this case, however, the information about $ M_{\rm min}$ is lost unless the isotropic component is subtracted.


\section{DM constraints}\label{sec:constrains}

\begin{figure}[t]
\includegraphics[angle=0.0, width=\columnwidth]{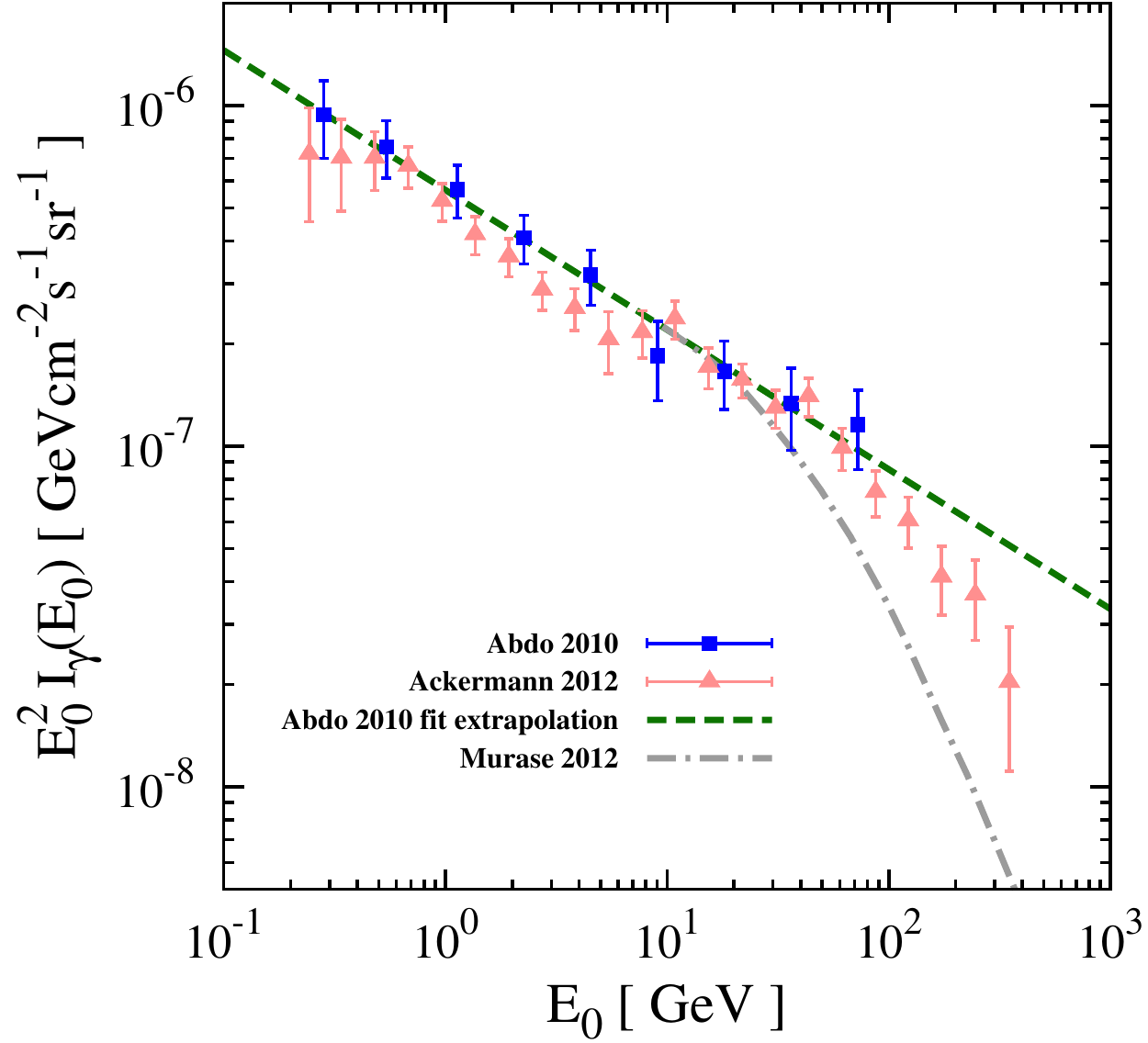}
\caption{
The IGRB spectrum measured by Fermi. We show the published Fermi IGRB data and the extended single power-law fit from Abdo et al.~\cite{Abdo:2010nz}, and also data points from Fermi preliminary results \cite{Ackermann:2013sy}. The EBL-attenuated power law is adapted from \cite{Murase:2012df}. 
 } 
\label{fig:plot_IGRB}
\end{figure}

In the previous section, we present the DM annihilation contribution to the IGRB from both the isotropic Galactic and extragalactic components. We consistently take into account the substructure enhancement using the K10 model. 

In this section, we show that comparing the signals from the GC and the IGRB can break the degeneracy of the small-scale cutoff ($ M_{\rm min}$) with the annihilation cross section, thus testing both cosmology and particle physics scenarios. We illustrate this using two DM candidate scenarios, representing the narrow line and the broad continuum classes. For simplicity, we focus the discussion of the clumping factor using the Halo Model approach with K10 substructure only. 

The energy spectrum of the IGRB measured by Fermi is shown in Fig.~\ref{fig:plot_IGRB}. We show the data points and the single power law fit, naively extrapolated, from the published Fermi result~\cite{Abdo:2010nz}, as well as preliminary result from Fermi \cite{Ackermann:2013sy}.  The attenuated power law is adapted from Murase et al.~\cite{Murase:2012df}, who considered the case that the IGRB is composed by unresolved astrophysical sources with star formation evolution. One can see the preliminary data set shows hints of spectral softening in high energies and is closer to the attenuated power law than the extrapolated power law. This could potentially lessen the Very High Energy Excess problem \cite{Murase:2012xs, Murase:2012df}, and it adds support to the hypothesis that the observed spectrum is extragalactic in origin, validating the background reduction procedure by Fermi. The attenuated power law represents one of the simplest astrophysical-only IGRB spectra, normalized to lower energy points. It shows the \emph{theoretical} limit of using the IGRB to constrain DM signals, if only flux information is used. In this work, we conservatively derive constraints using the extended power law fit, which is well above the curve with attenuation. 

\begin{figure}[t]
\includegraphics[angle=0.0, width=\columnwidth]{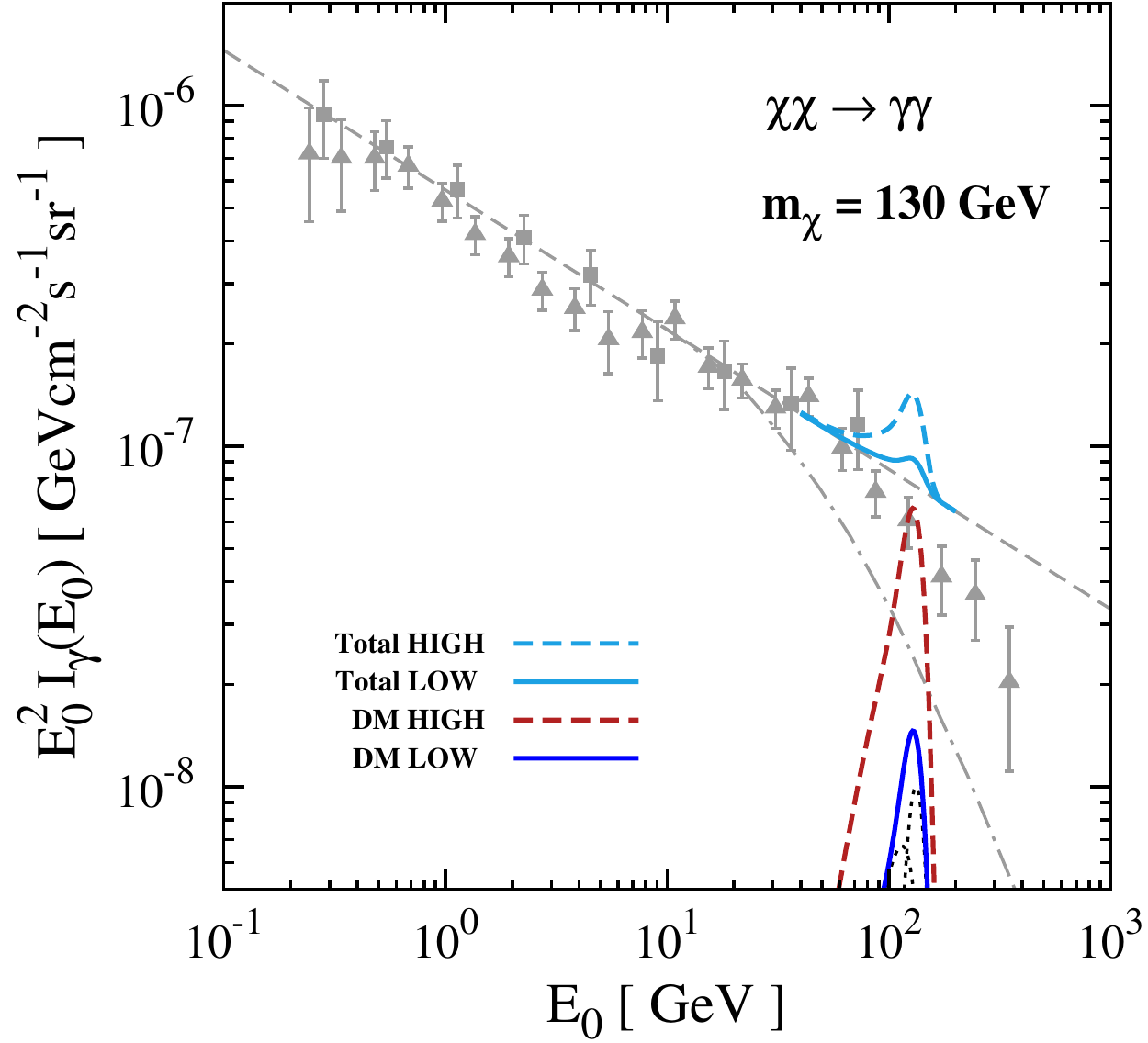}
\caption{
The combined (isotropic Galactic + extragalactic) DM signal for LOW and HIGH substructures for the 130 GeV DM with annihilation channel $\chi \chi \to \gamma \gamma$. Superposed are the IGRB spectra from Fig.~\ref{fig:plot_IGRB}. The individual isotropic Galactic and extragalactic components for the LOW substructure case are shown in black dotted lines. All DM components are evaluated with $\sigma v = {\rm 2\times 10^{-27}\,cm^{3}\,s^{-1}}$ and ${ M_{\rm min}} = 10^{-6}\,{ M_{\odot}}$. The DM signals are convolved with $10\%$ Gaussian smearing to take into account the energy resolution of Fermi-LAT. For visualization, we also show the total spectra (isotropic Galactic + extragalactic + Abdo 2010 fit) for LOW and HIGH cases.
 } 
\label{fig:plot_gamma}
\end{figure}

\begin{figure}[t]
\includegraphics[angle=0.0, width=\columnwidth]{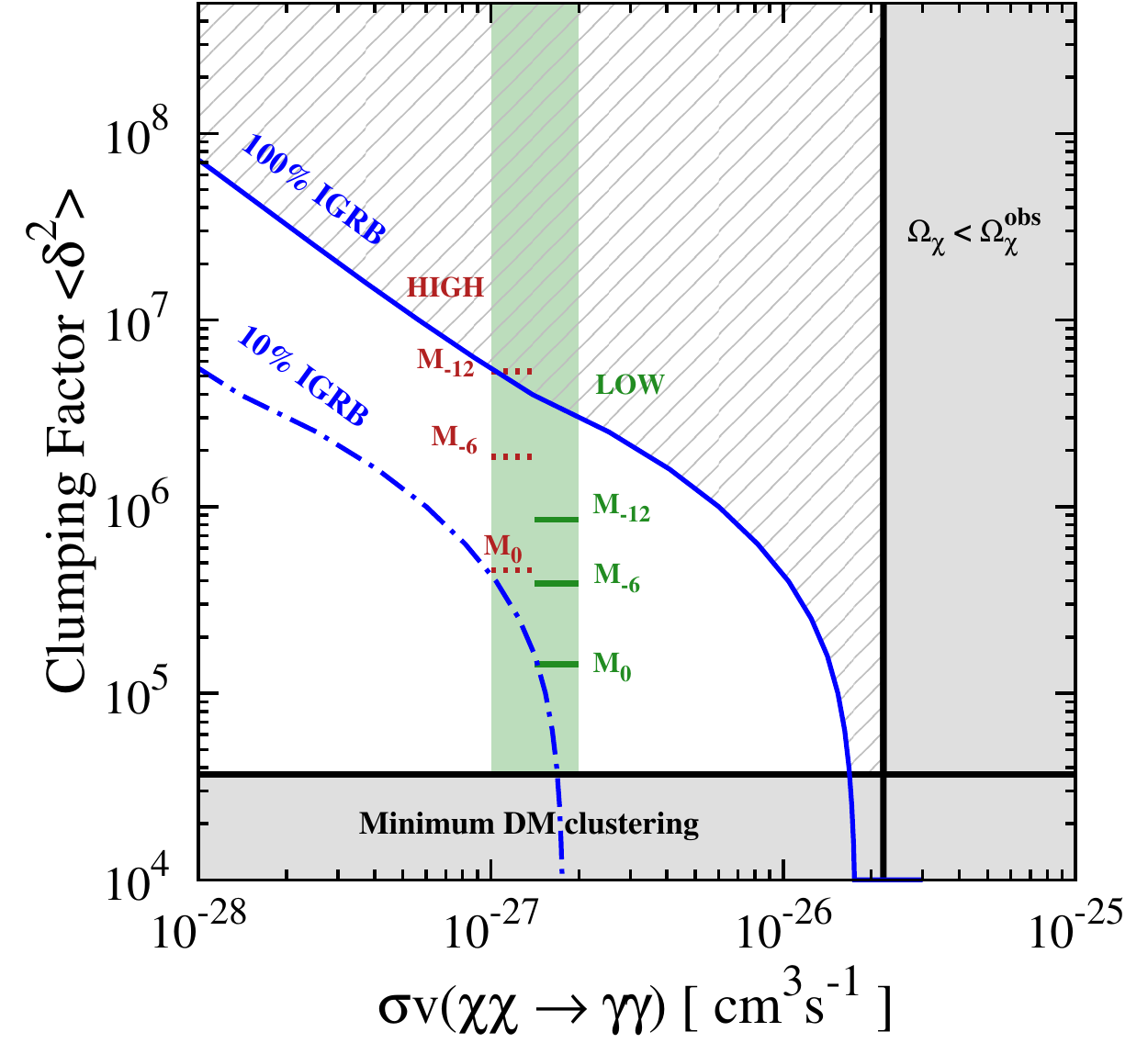}
\caption{
The ``clumping factor-$\sigma v$'' parameter space plane. It is bounded from below by the Minimum DM clustering and from the right by the relic abundance requirement. The allowed region for 130 GeV DM is fixed by the GC observation (vertical green band). The blue solid line is obtained using the total IGRB flux (100\% IGRB). The bending feature notes the transition into where the isotropic Galactic component dominates (Eq.~\ref{eq:schematic}). We show the translation from clumping factor to $ M_{\rm min}$ using the LOW (HIGH) substructure case, with solid green (dashed red) marks on the allowed parameter space. The blue dot-dashed line represents the parameter space that IGRB could probe for a more detailed analysis (10\% IGRB). } 
\label{fig:plot_gammacon}
\end{figure}

\subsection{Gamma-ray line -- 130 GeV DM} 

Recently, gamma-ray line with energy $\sim 130\,{\rm GeV}$ were reported towards the GC in Fermi-LAT data at high statistical significance, and clusters at less significance \cite{Bringmann:2012vr, Weniger:2012tx, Tempel:2012ey, Su:2012ft, Hektor:2012kc, Huang:2012yf}. So far, no such signals have been seen detected from dwarf galaxies \cite{GeringerSameth:2012sr}. There are astrophysical explanations for these events (\cite{Aharonian:2012cs}, but also see \cite{Carlson:2013vka}), as well as interesting instrumental effects \cite{Hektor:2012ev, Finkbeiner:2012ez,Whiteson:2013cs, Whiteson:2012hr}. Radio measurements of this candidate seems promising \cite{Asano:2012zv,Laha:2012fg} as a fairly model-independent check. The line signal, if interpreted as DM, requires the annihilation cross section to be $ \sigma v_{\chi\chi\rightarrow \gamma\gamma} = (1-2)\times 10^{-27} {\rm cm^{3}\,s^{-1}}$. This value of $\sigma v$ is higher than normally expected \cite{Bergstrom:1997fh}, but could be a manifestation of DM physics \cite{Tulin:2012uq}. The morphology of the signal is best fit with the Einasto profile, but is also consistent with the NFW profile \cite{Bringmann:2012ez}. The Fermi collaboration has confirmed the feature, but with lower significance and a small shift in energy to 133 GeV \cite{Fermi-LAT:2013uma}, mostly due to a better modeling of the detector response to monochromatic photons. The nature of this feature is currently inconclusive, and actions are advocated to quickly resolve the situation \cite{Weniger:2013tza}. In this section we assume the feature is due to DM annihilation, and refer to it as the 130 GeV DM.

One can predict the contribution to the IGRB from such a DM particle given the information from GC. We show a representative case in Fig.~\ref{fig:plot_gamma}. Assuming $ M_{\rm min} = 10^{-6}\, { M_{\odot}}$ and $\sigma v = 2 \times 10^{-27}\, {\rm cm^{3}\,s^{-1}}$, we show the combined isotropic Galactic and extragalactic DM components for both LOW and HIGH substructure cases. For the extragalactic component, we integrate up to redshift 4 to cover the interesting energy range. Both features are obtained by convolving the intrinsic spectrum using $10\%$ energy resolution with Gaussian smearing.

\begin{figure}[t]
\includegraphics[angle=0.0, width=\columnwidth]{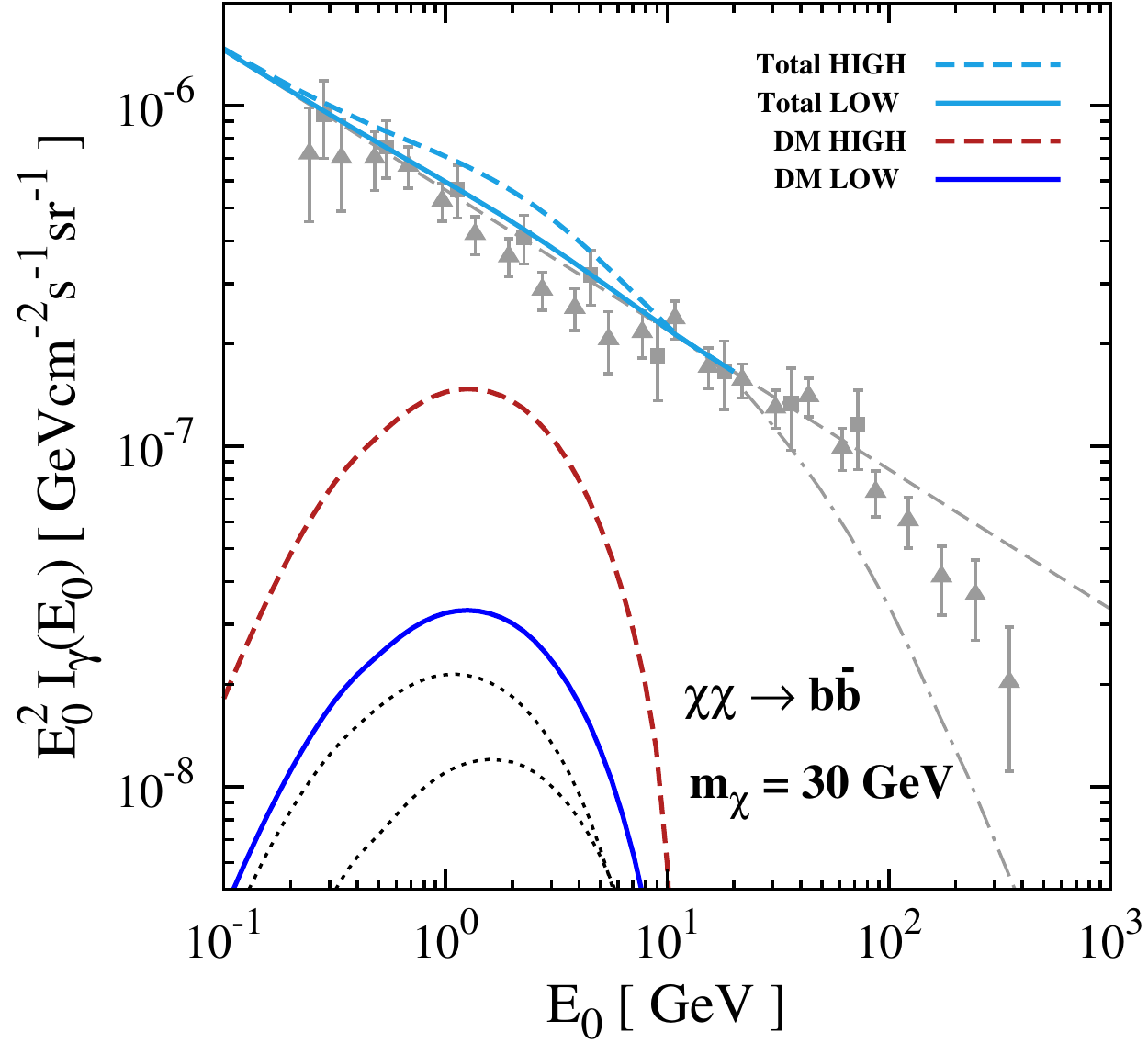}
\caption{Same as Fig.~\ref{fig:plot_gamma}, but with 30 GeV $\chi\chi\to b\bar{b}$ and $\sigma v = {\rm 2.2\times 10^{-26}\,cm^{3}\,s^{-1}}$. We consider the prompt photon spectrum only.
}
\label{fig:plot_b}
\end{figure}

We first consider the most conservative constraint of DM annihilation from the IGRB. This can be obtained by requiring the total DM signal to not overshoot the total flux of the IGRB. Recall that for the extragalactic component, the clumping factor is degenerate with the annihilation cross section. In addition, the clumping factor correlates with the Galactic substructure boost through their dependence on $ M_{\rm min}$. Therefore, the general constraint is a surface on the ${ M_{\rm min}}$-$\sigma v$-$m_{\chi}$ space. For a specific DM case, like the 130 GeV DM, we can condense the $m_{\chi}$ dimension. Lastly, the resultant ${ M_{\rm min}}$-$\sigma v$ plot would depend on the underlying model of the DM density distribution. A more convenient treatment is to construct the clumping factor versus $\sigma v$ plot. In that case, most of the model dependence moves to the interpretations of the parameter space. For a pure extragalactic component, a constant flux would be represented by a straight diagonal line in the clumping factor-$\sigma v$ plane, representing complete degeneracy. 

In Fig.~\ref{fig:plot_gammacon}, we show one of the main results of this work. The observed IGRB flux defines a line in the clumping factor-$\sigma v$ plane, as labeled by the 100\% IGRB line, above which the DM signal exceeds the IGRB total flux, and thus is robustly excluded. Superposed are two independent constraints. The plane is bounded from below by minimal DM clustering, conservatively defined by $ M_{\rm min} = 10^{6}\,{ M_{\odot}}$, and bounded from the right by the relic abundance criterion (the precise value of $\langle \sigma v \rangle$ is mass dependent and is $\sim 2.2 \times 10^{-26}\,{\rm cm^{3}\,s^{-1}}$ \cite{Steigman:2012nb}, for the mass range that we are interested in).

The degeneracy between the clumping factor and $\sigma v$ is apparent in the parameter space where the extragalactic component dominates the isotropic Galactic component. As one increases the value of $ M_{\rm min}$, the decrease of the clumping factor is much faster than the decrease of the boost factor for the isotropic Galactic component. When $\langle \delta^{2} \rangle$ falls below $\sim 8\times 10^{4}$, the Galactic component begins to dominate the extragalactic component (Eq.~\ref{eq:schematic}), resulting in near-independence of the flux on the value of $\langle \delta^{2} \rangle$, and hence the bending feature.  The required value of $\sigma v$ for the 130 GeV DM is labelled by the vertical green band. On the green band, we map the clumping factor to $ M_{\rm min}$ using the LOW substructure model (solid green labels). In this conservative scenario, the constraint is below $ 10^{-12}\, M_{\odot}$, and thus can be considered as unconstrained. But for the HIGH case (dotted red labels), it is probing near $ 10^{-12}\, M_{\odot}$. For simplicity, we conservatively ignored the larger enhancement for the isotropic Galactic component for the HIGH substructure case, which matters only when the isotropic Galactic component dominates. 

\begin{figure}[t]
\includegraphics[angle=0.0, width=\columnwidth]{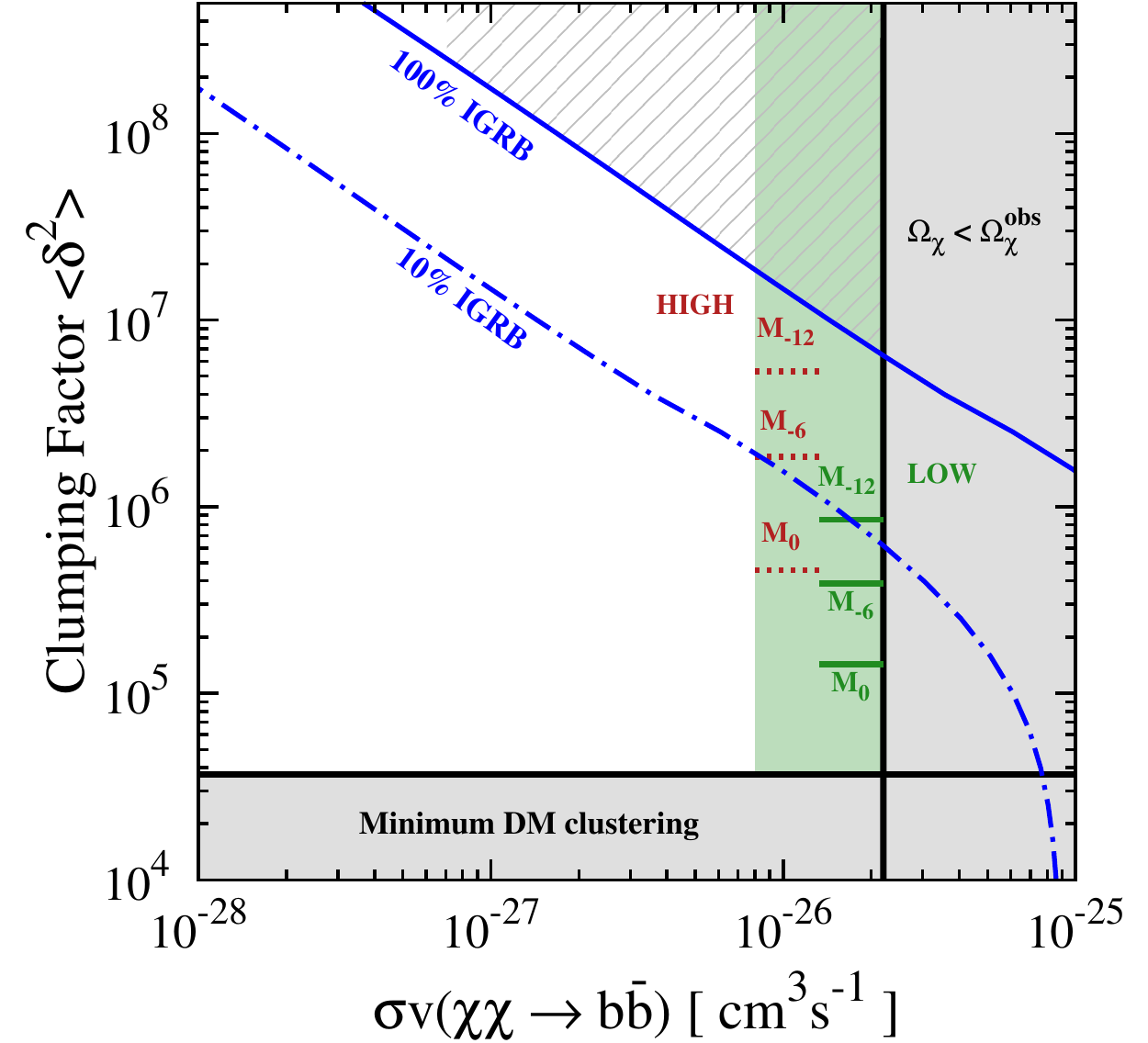}
\caption{Same as Fig.~\ref{fig:plot_gammacon}, but with 30 GeV  $\chi\chi\to b\bar{b}$. The green parameter space are taken from the ``best-fit spatial model'' from \cite{Abazajian:2012pn}. 
} 
\label{fig:plot_bcon}
\end{figure}

Any additional non-DM component will significantly improve the constraint. This extra component could come from different unresolved astrophysical sources, including star-forming galaxies, blazers, etc (e.g., see \cite{Dermer:2007fg, Dermer:2012ry}). 

One can also distinguish DM signals from the data itself. DM annihilation signals usually show a sharp spectral cut-off near the DM mass. Such features, if present in the data, should be easily isolated from any underlying background that behaves like power laws. The distinct anisotropy feature of DM annihilation is also a powerful tool to distinguish the DM signal from non-DM components, even down to $\sim 10\%$ level \cite{Ando:2009fp}.

Therefore, IGRB DM sensitivity can potentially reach 10\% of its total flux using either better background estimation, spectral analysis, or anisotropy analysis. We label this by the 10\% IGRB line in Fig.~\ref{fig:plot_gammacon}, above which is the parameter space we think realistic IGRB analysis can probe. One can see that with 10\% DM sensitivity, IGRB can probe up to $\sim 1 \,{ M_{\odot}}$, the upper extreme for most of the cold DM scenarios.  

In addition, we have conservatively taken the isotropic Galactic component to be from the Anti-GC. The IGRB analysis, however only uses photons from $\sim 80 \%$ of the sky ($|b|>10^{\circ}$) \cite{Abdo:2010nz}. A realistic estimation of how much the Galactic Halo DM component is contaminating the IGRB probably requires a detailed study by adding a DM template to the IGRB analysis. To estimate that analysis, we consider using the Galactic Pole (${\cal J}(\frac{\pi}{2})$), where the sky is least contaminated by the Galactic foreground. In this case, the isotropic Galactic component constraint, represented by the bending features in Fig.~\ref{fig:plot_gammacon}, is improved by a factor of $\sim2$ (the constraints lines shift to the left by a factor of 2). All of the 130 GeV DM parameter space will be probed by the 10\% IGRB line in this case. 

\begin{figure}[t]
\includegraphics[angle=0.0, width=\columnwidth]{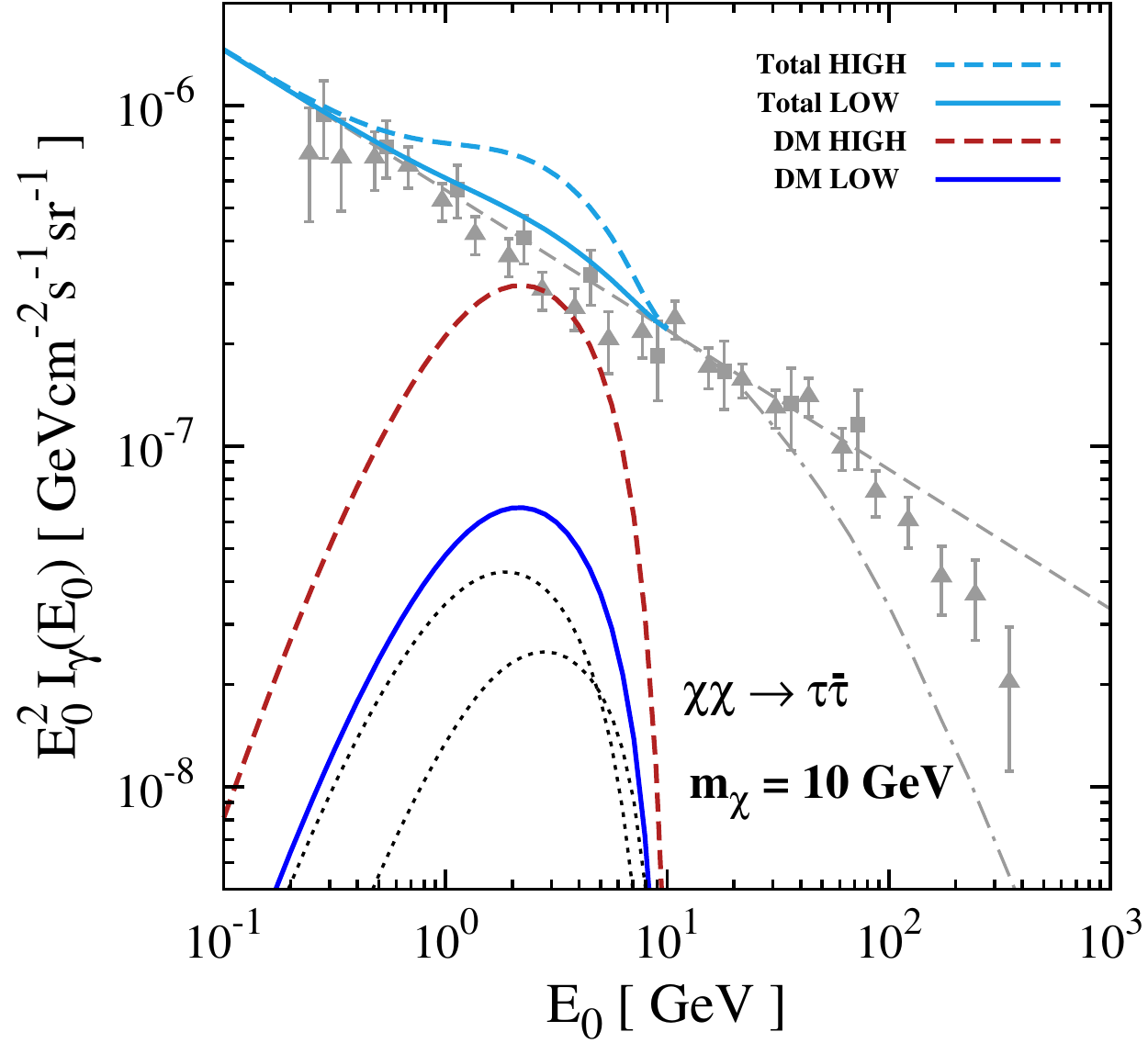}
\caption{Same as Fig.~\ref{fig:plot_gamma}, but with 10 GeV $\chi\chi\to \tau^{+}\tau^{-}$ and $\sigma v = {\rm 2.2\times 10^{-26}\,cm^{3}\,s^{-1}}$. We consider the prompt spectrum only.
}

\label{fig:plot_tau}
\end{figure}

\begin{figure}[t]
\includegraphics[angle=0.0, width=\columnwidth]{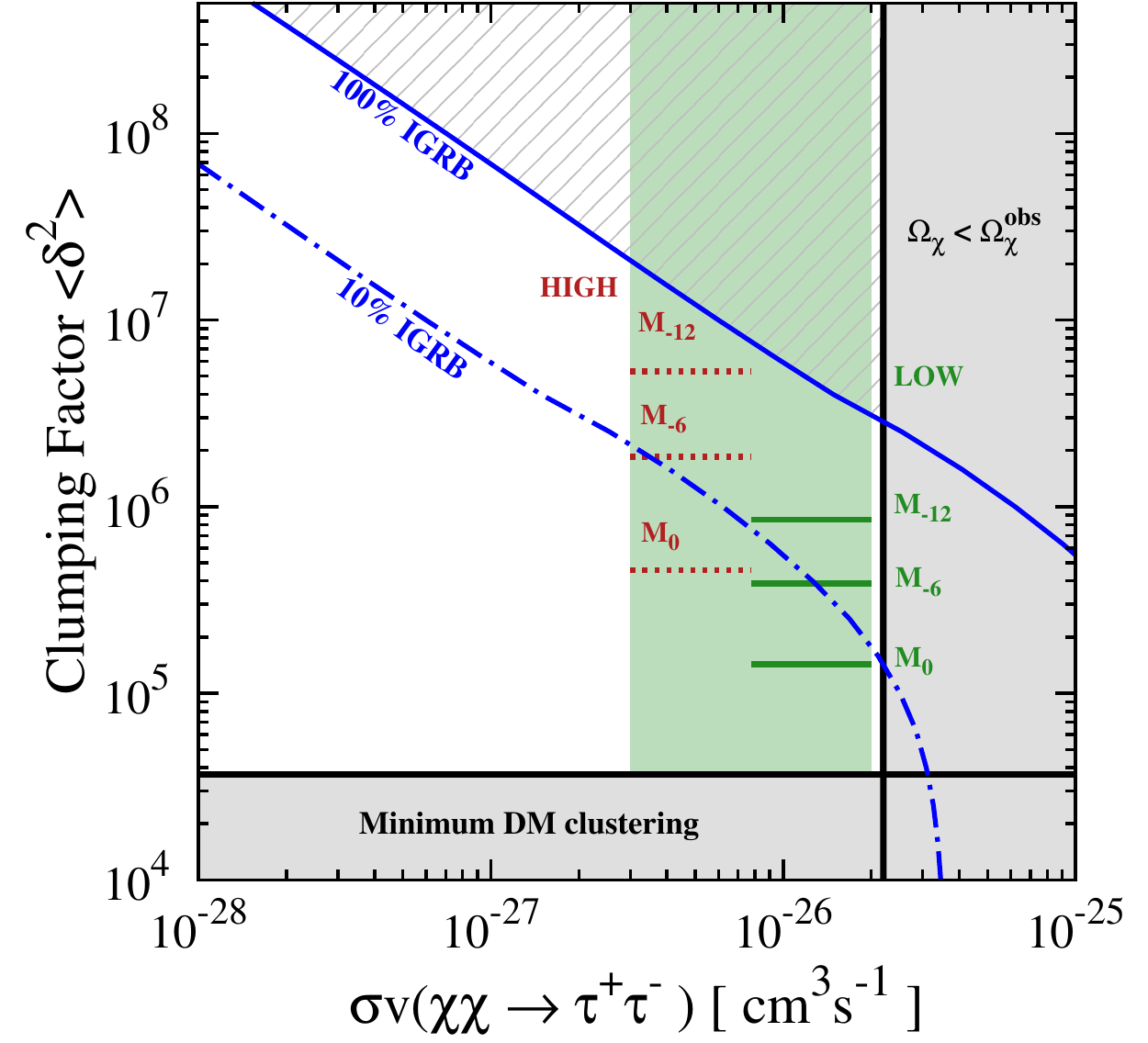}
\caption{Same as Fig.~\ref{fig:plot_gammacon}, but with 10 GeV  $\chi\chi\to \tau^{+}\tau^{-}$. The green parameter space are taken from the ``best-fit spatial model'' from \cite{Abazajian:2012pn}.} 
\label{fig:plot_taucon}
\end{figure}

Last but not least, we used the spectral fit to the Fermi data given by Ref.~\cite{Abdo:2010nz} to derive all the constraints. With more and better data \cite{Atwood:2013rka}, as already partly demonstrated by Ref.~\cite{Ackermann:2013sy}, one can expect the overall constraint can be improved significantly soon, especially if the high energy spectral softening is confirmed. 

We therefore conclude that a DM IGRB analysis in the near future can realistically probe all of the parameter space of the 130 GeV DM, even in the conservative substructure case. Such an analysis contains slightly different systematics than the GC DM search, since the GC region is excluded from the IGRB measurement.

\subsection{Continuum -- 10-30 GeV light DM} 
For many DM models, DM annihilating into quarks or leptons is more favored than monochromatic photons, since the latter may be loop-suppressed. Therefore, annihilations typically produce a broad gamma-ray continuum. Much attention has been paid to the low energy spectrum observed by the Fermi-LAT at the GC where unexplained excess photons are observed \cite{Goodenough:2009gk,Hooper:2010mq, Hooper:2010im, Hooper:2011ti, Abazajian:2012pn, Hooper:2012sr, Hooper:2013rwa}, which may be incompatible with being unresolved astrophysical sources \cite{Hooper:2013nhl} (but also see \cite{Mirabal:2013rba}). To obtain the GC excess, the complicated GC astrophysical emission needs to be subtracted. The resulting excess is therefore subject to large systematic uncertainties.

If interpreted as signals from DM annihilation, these excesses are generally favored by $\chi\chi \to b\bar{b}$ or $\chi\chi \to \tau^{+}\tau^{-}$ at $\sigma v ~{\sim 10^{-26}\, \rm cm^{3}\,s^{-1}}$ and mass $10-30$ GeV. The profiles favored by the excesses are usually more cuspy than NFW (typically $\gamma \sim 1.3$). The cuspy profile has no impact on the calculation of the isotropic Galactic component. It can constitute an extra boost to the clumping factor. We conservatively continue considering the NFW case for the extragalactic component.

Similar to the 130 GeV DM, we test the compatibility of these DM annihilation channels using the IGRB. For definiteness, we use the best fit parameters from Ref.~\cite{Abazajian:2012pn}, which are $\rm m_{\chi}=30\,GeV$ for the $b\bar{b}$ and $\rm m_{\chi}=10\,GeV$ for $\tau^{+}\tau^{-}$ channels, respectively. We consider a range of $\sigma v$ with $(0.8-2.2)\times 10^{-26}\, {\rm cm^{3}\,s^{-1}}$ for $b\bar{b}$ and $(0.3-2)\times 10^{-26}\, {\rm cm^{3}\,s^{-1}}$ for $\tau^{+}\tau^{-}$ given by the best-fit spatial model ($\gamma = 1.2$) in the above reference. 

In Fig.~\ref{fig:plot_b} and \ref{fig:plot_tau}, we show the spectra of $\chi\chi \rightarrow b\bar{b}$ and $\chi\chi \to \tau^{+}\tau^{-}$ together with the IGRB data, in the same format as Fig.~\ref{fig:plot_gamma}. We adopt $\sigma v = 2.2\times 10^{-26}\, {\rm cm^{3}\,s^{-1}}$ and ${ M_{\rm min}} = 10^{-6}\,{ M_{\odot}}$. The gamma-ray spectra are obtained using Pythia \cite{Sjostrand:2007gs}. 

In Fig.~\ref{fig:plot_bcon} and \ref{fig:plot_taucon}, we show the corresponding constraints in the clumping factor-$\sigma v$ plane. The $\tau^{+}\tau^{-}$ channel constraint is slightly better than the $b\bar{b}$ due to the slightly smaller DM mass and harder spectrum. The conservative constraint, by requiring the DM signals do not overshoot the total IGRB flux, is given by the 100\% IGRB line. Using the conservative LOW substructure case, we see that both $b\bar{b}$ and $\tau^{+}\tau^{-}$ are unconstrained. For the HIGH substructure case, the 100\% IGRB line would carve into parameter space near $10^{-12}\,{ M_{\odot}}$, for the $\tau^{+}\tau^{-}$ channel. 

Similar to the 130 GeV DM case, any extra component in the IGRB or any method in isolating the potential DM signal from background can significantly shrink the allowed parameter space. The spectra of $b\bar{b}$ and $\tau^{+}\tau^{-}$ are not as sharp as the monochromatic photon channel, but they do have a cutoff in the spectrum near the DM mass. The light DM annihilation channels also enjoy higher statistics compared to the 130 GeV DM, which would benefit both the spectral and anisotropy analyses. We therefore also consider $10\%$ DM sensitivity as realistic for the light DM, as shown by the 10\% IGRB line in Fig.~\ref{fig:plot_bcon} and \ref{fig:plot_taucon}. In that case, even for the conservative LOW substructure case, the IGRB can probe near $10^{-6}\,{ M_{\odot}}$, and even up to $10^{0}\,{ M_{\odot}}$ for higher $\sigma v $ regions. As a result, the IGRB is also promising in constraining optimistic substructure and small $ M_{\rm min}$ DM scenarios. 

For both channels, we only considered the prompt photon emission and ignored secondary processes such as synchrotron, bremsstrahlung and inverse Compton emissions. These components depend on the astrophysical environments such as photon density and magnetic field. Adding these components would improve the constraints on the specific channels. For more thorough treatments of these processes, see, e.g., Ref.~\cite{Abazajian:2010zb, Blanchet:2012vq,Cirelli:2013mqa}.  The DM candidates we considered, were obtained from fits without taking into account secondary emissions. As a result, for easier comparison, we neglect these extra components. 

\section{Summary and Outlook}
\label{sec:summary}
\subsection{Summary}
We study the effect of Dark Matter (DM) microhalos on DM annihilation signals in the Isotropic Gamma-Ray Background (IGRB). We demonstrate that using substructure-dominated systems and multi-source observations together can constrain the minimum halo mass and annihilation cross section separately. We show that using the IGRB leads to interesting sensitivity for testing tentative signals from the Galactic Center (GC).

We consider the case of DM annihilation contributing to the IGRB. Motivated by Prada et al.~(\cite{Prada:2011jf}, P12), we extend their results using a physically-motivated cosmological variable, $\sigma_{L}(M,z)$, with the latest Planck cosmology. As a result, we obtain a halo concentration description that fits well to both large-scale observations and small-scale microhalo simulations. Adding the substructure model of Kamionkowski, Koushiappas, and Kuhlen \cite{Kamionkowski:2010mi}, we consistently take into account the effect of DM substructures on the isotropic Galactic and extragalactic signals of DM annihilation. For a given substructure scenario, the IGRB DM contribution then only depends on the minimum halo mass, $ M_{\rm min}$, set during the epoch of kinetic decoupling, and the annihilation cross section, $\sigma v$. 

We show that using the IGRB alone, the DM constraint suffers from the ``clumping factor-$\sigma v$'' degeneracy. We propose a new perspective by constructing the ``clumping factor-$\sigma v$'' plots, where this problem is explicit for any particular DM case (Figs.~\ref{fig:plot_gammacon}, \ref{fig:plot_bcon}, \ref{fig:plot_taucon}). The degeneracy can be broken by adding information from an independent measurement, thus yielding information for both $ M_{\rm min}$ and $\sigma v$. This is potentially the only method to observationally constraint $ M_{\rm min}$ for cold DM cosmologies. 

We demonstrate this idea by comparing the Fermi-measured IGRB to two tantalizing DM gamma-ray indirect detection candidates from the GC. One is the $\sim$ 130 GeV DM in the $\chi\chi \to \gamma\gamma$ channel. The other is ($10-30$) GeV light DM in the $\chi\chi \to b\bar{b}$ or $\chi \chi \to \tau^{+}\tau^{-} $ channels. We show that, in the most conservative case, where the substructure fraction is low and DM annihilation is allowed to saturate the IGRB flux, DM analyses using the IGRB are reaching interesting sensitivity for $ M_{\rm min}$.

We further argue that it is unlikely that DM annihilation signals would dominate the IGRB. Taking into account unresolved astrophysical sources can reduce the allowed DM contribution to the IGRB. Utilizing the spectral and anisotropy feature of DM annihilation signals, one could further limit the IGRB DM component. We show that if  10\% DM sensitivity can be achieved by a more detailed analysis using the IGRB, one should be able to recover the 130 GeV DM signal, while the more clumpy cases can be probed for light DM. The rapid improvement of the limit on $M_{\rm min}$ reflects the physical expectation that concentrations increase progressively slower with decreasing scales, as shown by the P12 concentration-mass relation.  Last but not least, we use the single power law fit from the Fermi published result, which only uses 2 years of data. Imminent Fermi updates on the IGRB with better data in terms of background reduction and higher statistics would further improve our result. 

\subsection{Outlook}

We only focus on the velocity independent DM annihilation case. One can generalize and probe the velocity dependent DM candidates (e.g.,~\cite{Hisano:2004ds, ArkaniHamed:2008qn}). In that case, in addition to the DM spatial distribution, one can probe the DM velocity distribution as well. The relevant quantity for the extragalactic component would be $\langle \rho_{\chi}^{2}\sigma v \rangle$ \cite{Campbell:2010xc, Campbell:2011kf}. This is analogous to the clumping factor, but also takes into account the velocity distribution. 

We demonstrate the benefits of comparing GC and IGRB for DM annihilation signals, but one need not stop there. In principle, one can do global analyses including multiple DM sources, e.g., observations from Dwarf Galaxies or Galaxy Clusters etc.  It can further disentangle different dependencies like halo profiles, substructure content, substructure evolution history, etc. 

We have reached the era where many astrophysical probes are reaching the relevant parameter space for simple WIMP DM indirect detections. We anticipate that in the future cross correlation of multiple astrophysical observations will become more and more important in DM indirect detection.

\bigskip
\noindent
{\bf Acknowledgments:}
We are grateful to Paolo Gondolo, Savvas Koushiappas, Shirley Li, and Miguel S\'anchez-Conde for helpful comments. KCYN, RL, SC, and JFB were supported by NSF Grant PHY-1101216 to JFB, and BD, SH, and KM were supported by CCAPP during the early part of this work. SH was supported by a JSPS fellowship for research abroad. KM was supported by NASA through Hubble Fellowship, Grant No. 51310.01 awarded by the STScI, which is operated by the Association of Universities for Research in Astronomy, Inc., for NASA, under Contract No. NAS 5-26555.

\bibliographystyle{h-physrev}
\bibliography{Bibliography/references}

\end{document}